\documentclass[11pt,preprint]{aastex}

\newcommand{\kms}{km~s$^{-1}$}
\newcommand{\oii}{[O{\scshape ii}]}
\newcommand{\oiii}{[O{\scshape iii}]}
\newcommand{\nii}{[N{\scshape ii}]}
\newcommand{\sii}{[S{\scshape ii}]}
\newcommand{\ha}{H$\alpha$}
\newcommand{\hb}{H$\beta$}
\newcommand{\hd}{H$\delta$}

\begin{document}

\title{Abell 2111:\\
An Optical and Radio Study\\
of the Richest Butcher-Oemler Cluster}

\author{Neal A. Miller\altaffilmark{1,2}} 
\email{nmiller@pha.jhu.edu}

\author{William R. Oegerle\altaffilmark{3}}

\author{John M. Hill\altaffilmark{4}}

\altaffiltext{1}{Jansky Fellow of the National Radio Astronomy Observatory. The National Radio Astronomy Observatory is a facility of the National Science Foundation operated under cooperative agreement by Associated Universities, Inc.}
\altaffiltext{2}{Department of Physics and Astronomy, Johns Hopkins University, 3400 N. Charles Street, Baltimore, MD 21218}
\altaffiltext{3}{NASA Goddard Space Flight Center, Exploration of the Universe Division, Code 660, Greenbelt, MD 20771}
\altaffiltext{4}{Large Binocular Telescope Observatory, 933 N. Cherry Avenue, Tucson, AZ 85721-0065}

\begin{abstract} 
We present an in-depth analysis of the Butcher-Oemler cluster A2111, including new optical spectroscopy plus a deep Very Large Array (VLA) radio continuum observation. These are combined with optical imaging from the Sloan Digital Sky Survey (SDSS) to assess the activity and properties of member galaxies. Prior X-ray studies have suggested A2111 is a head-on cluster merger, a dynamical state which might be connected to the high level of activity inferred from its blue fraction. We are able to directly assess this claim, using our spectroscopic data to identify 95 cluster members among 196 total galaxy spectra. These galaxy velocities do not themselves provide significant evidence for the merger interpretation, however they are consistent with it provided the system is viewed near the time of core passage and at a viewing angle $\gtrsim$30\degr{} different from the merger axis. The SDSS data allow us to confirm the high blue fraction for A2111, $f_b=0.15\pm0.03$ based on photometry alone and $f_b=0.23\pm0.03$ using spectroscopic data to remove background galaxies. We are able to detect 175 optical sources from the SDSS in our VLA radio data, of which 35 have redshift information. We use the SDSS photometry to determine photometric redshifts for the remaining 140 radio-optical sources. In total we identify up to 26 cluster radio galaxies, 14 of which have spectroscopic redshifts. The optical spectroscopy and radio data reveal a substantial population of dusty starbursts within the cluster. The high blue fraction and prevalence of star formation is consistent with the hypothesis that dynamically-active clusters are associated with more active member galaxies than relaxed clusters.\end{abstract}

\section{Introduction}

The discovery of the Butcher-Oemler effect \citep[][hereafter BO84]{butc1984} has led to an unusual bias in the study of galaxy evolution in clusters. As the Butcher-Oemler effect indicates galaxy clusters were more active at earlier epochs, there has been great interest in the study of clusters at higher and higher redshift. This interest has also been fueled by better observational capabilities, which have increasingly made it possible to obtain high quality data at fainter limits. These same factors have sparked continued interest in low redshift clusters of galaxies; well-studied low redshift clusters are the necessary benchmark for comparison with higher redshift samples, and provide the best opportunity to probe deep into luminosity functions. The result of these trends is that many intermediate redshift clusters ($z \sim 0.2$) have been largely overlooked.

Abell 2111, at $z=0.23$, has the distinction of being the richest cluster in the original BO84 study with $N_{30} = 155$ (i.e., 30\% of the cluster's total projected galaxies amounts to 155 galaxies). For reference, the next richest cluster in that study, A520, had $N_{30} = 126$ and the 3C295 cluster which inspired the original research had only $N_{30} = 45$. A2111 also was among the higher blue fraction clusters noted in BO84 at $f_b = 0.16 \pm 0.03$. The primary concern with the interpretation of the BO84 results was that of contamination, prompting subsequent campaigns to obtain optical spectroscopy and confirm whether the blue galaxies were truly cluster members or simply galaxies viewed in projection on the clusters. Consequently, spectroscopy results for 24 galaxies in the direction of A2111 were presented by \citet{lavery1986}. These included 14 cluster members and confirmed the high blue fraction determined photometrically by BO84. The same authors further investigated the Butcher-Oemler effect using their A2111 data along with several other clusters, and found evidence that the blue galaxies often had nearby companions indicative of galaxy-galaxy mergers \citep{lavery1988} and that their morphologies frequently included disks \citep{lavery1994}. However, outside of these papers and the presentation of galaxy morphologies within the very core of the cluster \citep{fasano2000}, there has been little optical study of A2111.

In addition to its richness and high blue fraction, X-ray studies of A2111 reveal an interesting cluster. \citet{wang1997} presented {\sl ROSAT} observations of A2111 in conjunction with optical imaging in $R$-band. The X-ray data revealed a complex morphology for the intracluster gas, including elongated emission whose degree of elongation changes with intensity. The cluster X-ray emission was further decomposed into a main component with a superimposed subcomponent. The subcomponent consisted of a main central blob located between the two bright galaxies at the cluster center, a diffuse twisted tail to the northwest, and a lobe to the southeast. This was interpreted as evidence for A2111 representing a head-on merger. The subcluster would have entered from the northwest, leaving the diffuse tail. Its core, the central blob of emission, would be heated as it plowed through the center of the main cluster component. The {\sl ROSAT} data did indicate a higher temperature for the blob, which was later confirmed by {\sl ASCA} \citep{henriksen1999}. Related studies have argued that cluster mergers evidenced by the elliptical appearance of their X-ray emitting intracluster gas are intimately related to galaxy evolution and high blue fractions \citep{wang1997b,metevier2000}.

Recent investigations of the radio galaxy populations of clusters have also yielded tantalizing evidence that cluster mergers may be associated with increased galaxy activity. The best-studied example is A2125, a high blue fraction cluster at $z=0.25$. In an initial study using the NRAO Very Large Array (VLA), \citet{dwarka1999} and \citet{owen1999} found a much higher detection rate for A2125 galaxies relative to the comparable richness but low blue fraction cluster A2645. The significance of the radio data is that radio emission from individual galaxies is an excellent tracer of star formation and AGN activity \citep[e.g.,][]{condon1992,yun2001}. In the absence of an AGN, the radio continuum emission (typically at 1.4~GHz) is a nearly direct measure of the star formation rate (SFR). The principal difference between A2125 and A2645 was their dynamical states, with A2125 being a probable cluster merger and A2645 a relaxed cluster. Further investigation of A2125, including optical spectroscopy of hundreds of galaxies \citep{my2125} and deeper, higher resolution radio observations and ancillary data \citep{owen2005} supported the dynamical assessment and strengthened the association of the radio detections with actively star-forming galaxies. Additional examples of dynamically-active clusters with increased populations of radio galaxies are gradually being identified \citep[A2255, A3562, A851;][]{my2255,my3562,morrison2005}. In the case of A851, radio observations proved especially fruitful as several galaxies with post-starburst optical spectra -- and hence no optical evidence for current star formation -- appear to be heavily dust-enshrouded starbursts \citep{smail1999}. The activity of these galaxies was revealed through their detection at radio wavelengths.

In this paper, we present a detailed multiwavelength analysis of A2111. It includes much more comprehensive optical spectroscopy than previously published, with over 200 collected spectra. These data enable us to present a dynamical analysis of the cluster in order to assess the hypothesis that A2111 is a cluster merger. The Sloan Digital Sky Survey \citep[SDSS,][]{york2000} has also recently covered A2111, thereby providing accurate multiband photometry of cluster galaxies. With these data and our spectroscopy, we re-investigate the cluster blue fraction: is A2111 truly a ``blue'' cluster, or is it significantly contaminated by foreground and background galaxies? What do the spectra tell us about the cluster galaxies -- is there evidence for current or recent starbursts, potentially related to the cluster dynamical state? Finally, we present a deep VLA radio continuum observation of A2111. The radio data achieve an rms sensitivity of 13 $\mu$Jy, thereby detecting cluster galaxies with SFR as low as 5.5 M$_\odot$ year$^{-1}$ \citep[using the relationship of][which assumed a Salpeter IMF from 0.1 to 100 M$_\odot$]{yun2001}. In addition to study of the cluster itself, the multiwavelength data are a useful dataset for understanding galaxy evolution from deep field observations. Thus, all optical sources with associated radio detections are reported along with useful information such as photometric redshifts (or spectroscopic redshifts when available).

Unless otherwise stated, we adopt the {\it Wilkinson Microwave Anisotropy Probe} cosmology with $H_0 = 71$ \kms{} Mpc$^{-1}$ and $\Omega_m =0.27$ and $\Omega_\Lambda=0.73$ \citep{spergel2003}. This translates to 1\arcsec{} equaling 3.6 kpc at A2111. The assumed cluster center position is 15$^{\mbox{{\scriptsize h}}}$39$^{\mbox{{\scriptsize m}}}$40\fs9  $+$34\degr25\arcmin04\arcsec{} (all reported coordinates are J2000) taken from the X-ray analysis of \citet{wang1997}. This lies between the two brightest optical galaxies, often associated with the cluster center. The first of these, at 15$^{\mbox{{\scriptsize h}}}$39$^{\mbox{{\scriptsize m}}}$40\fs5  $+$34\degr25\arcmin27\arcsec, has historically been used as the cluster center \citep[e.g.,][]{sand1976}. The second is located 47\arcsec{} away at 15$^{\mbox{{\scriptsize h}}}$39$^{\mbox{{\scriptsize m}}}$41\fs8  $+$34\degr24\arcmin43\arcsec.

\section{Data}

\subsection{Optical Spectroscopy}\label{sec-spec}

Galaxy spectra within the field of A2111 were obtained using the Hydra multifiber spectrograph on the WIYN 3.5 m telescope during 1996. Several clusters were the targets of these observations, and the reader is referred to \citet{my2125} for additional details on the data collection, reduction, and analysis. Briefly, three fiber configurations were observed during 1996 May 20--22, with a fourth observed on 1996 June 10--11 (this fourth configuration was obtained during queue mode scheduling). Typically around 30 of the 96 available fibers were assigned to random locations to produce a good quality sky spectrum for subtraction. The observations all used the blue fibers (3\farcs1) and a 400 line mm$^{-1}$ grating blazed at 4000$\mbox{\AA}$, with the resulting spectra having a resolution of 7$\mbox{\AA}$ and spanning from about 4300$\mbox{\AA}$ to 7500$\mbox{\AA}$. Each fiber configuration was observed multiple times to facilitate cosmic ray rejection, with total exposure times ranging from 2 to 5 hours. Standard calibration exposures were also obtained (bias frames, flat fields, fiber flats for tracing extraction apertures, and CuAr lamps for wavelength calibration), however, no flux calibration was performed. 

Velocities were measured by cross correlation with velocity standards using the IRAF RV package \citep{tonry1979}, and by Gaussian fitting of emission lines. For the cross correlations, a set of 22 velocity standard templates were used with the adopted velocity representing a weighted average of the results from the individual templates \citep[for details see][]{my2125}. Initial errors in cross correlation velocities were based on the goodness of fit, $R$, such that $\delta cz = 280(1 + R)^{-1}$ \kms{} \citep{hill1998}. A velocity for an emission line object was simply determined as the mean of the velocity implied by each of its fitted lines, shifted by the heliocentric correction. The initial error for an emission line measured velocity was simply the dispersion of the individual line fits divided by the square root of the number of fitted lines (i.e., the error in the mean). If a given galaxy had both a cross correlation and an emission line redshift, we adopted the measurement with the lower error. There were 20 such galaxies, and the velocities were found to be highly consistent with the largest difference between a cross correlation and emission line redshift being 120 \kms. Finally, an additional error of 40 \kms{} was added in quadrature to the error for each measured velocity. This value was determined based on repeated observations of velocity standards over the course of the observations \citep[see][]{my2125}.

The final velocities are presented in Table \ref{tbl-vels}. In addition, we have used the NASA/IPAC Extragalactic Database (NED) to identify galaxies with velocity measurements in the literature and within 22\farcm5 of the center of A2111, with this radius chosen based on our radio data (see Section \ref{sec-radio}). It corresponds to a linear diameter of 9.8 Mpc at the redshift of A2111. In total, there were 13 galaxies in the A2111 field which had NED velocities, six of which are also found in Table \ref{tbl-vels}. Our velocities are on average about 100 \kms{} greater, with a large dispersion of nearly 700 \kms. The NED velocities come from three prior works: \citet{lavery1994}\footnote{A prior paper \citep{lavery1986} detailed the spectroscopic data, which were collected in the mid-1980s, but the derived redshifts were not specifically noted until \citet{lavery1994}.}, \citet{crawford1999}, and \citet{morrison2003}. The bulk of the discrepancy between our velocities and those in NED can be attributed to a fairly large offset from the \citet{lavery1994} values, which are fairly coarse estimates (they are presented in terms of redshift out to three decimal places, and hence have an implied error of $\sim300$ \kms). For the three galaxies in common with Lavery \& Henry, our velocities are on average about 600 \kms{} greater. Similarly, spectroscopy covering the A2111 field has recently become available in the Sloan Digital Sky Survey Data Release 4 \citep[DR4,][]{sdssdr4}. There are 19 galaxy velocities in common with those in Table \ref{tbl-vels}. The mean difference in velocity between our measurements and the SDSS values is 15 \kms{} with a dispersion of 56 \kms, and hence an associated error in the mean of 13 \kms. Thus, our velocities are consistent with those of the SDSS. We also note this is a useful check on our derived errors. The dispersion in the mean velocity difference between our values and those of the SDSS represents the combined errors of the two measurements. The SDSS quotes an rms velocity error of $\sim$30 \kms, which would imply our velocity error is $\sim$45 \kms. Indeed, this uncertainty is typical of those velocities presented in Table \ref{tbl-vels}.

\subsection{Optical Imaging}\label{sec-image}

High quality optical imaging of A2111 is available from the SDSS Data Release 3 \citep[DR3][]{sdssdr3}. This provides imaging in five filters, $u~g~r~i~z$, with point source detection limits in magnitudes ranging from 20.5 ($z$) to 22.2 ($g$ and $r$) and photometric calibration good to about 2\%. The full width half maximum of the PSF in $r$ is typically around 1\farcs4.

Photometric data for all objects within the A2111 field were obtained from the DR3. The adopted magnitudes corresponded to the ``best'' magnitudes, which are determined using the latest version of the SDSS reduction pipeline \citep[the pipeline has been unchanged since the release of the DR2][]{sdssdr2}. They are model magnitudes based on the AB magnitude system, and correspond to the better of an exponential or DeVaucouleurs profile fit. In addition to this catalog data, FITS images were obtained for visual inspection of possible radio sources (see Section \ref{sec-ids}). We have also used an $R$ image kindly provided by Glenn Morrison for visual confirmations of sources in the inner regions of the surveyed area. This image was obtained using the KPNO 0.9m with the T2KA CCD, providing a 23\arcmin{} by 23\arcmin{} field of view. Details may be found in \citet{morrison2003}.

\subsection{Radio Continuum Imaging}\label{sec-radio}

\subsubsection{Observations and Reductions}

The radio data were collected using the VLA under program code AM759. A total of ten hours of time were scheduled, consisting of a pair of five-hour tracks centered on the transit of A2111 on 2003 June 4 and 2003 June 9. The VLA was in its largest configuration (A array), producing 1\farcs5 resolution images for the 1.4~GHz observations. The usual ``4'' mode was used for the correlator, which results in seven 3.125~MHz channels at each of two IFs for both right and left circular polarization. This allows for deep imaging by reducing the effects of bandwidth smearing over the large primary beam area. An integration time of 10 seconds was used, which reduced the size of the ({\it u,v}) data set at the expense of some time-average smearing. The pointing center was 15$^{\mbox{{\scriptsize h}}}$39$^{\mbox{{\scriptsize m}}}$41$^{\mbox{{\scriptsize s}}}$ $+$34\degr25\arcmin04\arcsec.

Calibration was achieved in AIPS via the standard procedures. The raw data were loaded including antenna-specific weights, which account for differences in observed system temperature among the 27 antennas. Flux calibration was based on observations of 3C286, and the source 1602+334 was observed roughly every 45 minutes for the purpose of phase and bandpass calibration. It lies within 5\degr{} of A2111. Additional self-calibration of the ({\it u,v}) data was performed during the imaging steps.

Imaging was performed using the AIPS task IMAGR. This included multiple facets to account for sky curvature over the large primary beam size (the ``3D effect''). The majority of the primary beam, a radius of 22\farcm5 which translates to about the radius where the response is about 20\% of that at the pointing center, was imaged in 37 facets each of 1024 $\times$ 1024 0\farcs4 pixels. An additional 19 facets were devoted to bright sources at the very edges of the primary beam and in the sidelobes. The overall procedure was incremental, with imaging followed by inspection of the images and boxing of detected sources to be used as input to subsequent imaging runs. Self-calibration steps were also included, initially just in phase and later in both amplitude and phase. The final images had a 1\farcs51 $\times$ 1\farcs50 FWHM beam, with most facets having an rms noise of 13 $\mu$Jy. The noise was higher in the vicinity of a few bright sources, although only one of these was located within 2 Mpc of the cluster center (J153912.0+342933). Near this source, the rms noise rises to 20 $\mu$Jy.

A catalog of radio sources was then constructed from the 37 primary facets using the AIPS task SAD. This catalog was meant to be fairly generous in detection of sources, and SAD extracted sources for which the peak and integral fluxes were above 52 $\mu$Jy (i.e., 4$\sigma$ sources at the pointing center). SAD was also directed to create residual maps which were inspected to identify any sources missed by the algorithm. These were added to the output catalog manually. All possible sources were later subjected to more rigorous standards after correlating with the optical data, as detailed in the next section. 

\subsubsection{Optical IDs for Radio Sources}\label{sec-ids}

\begin{figure}
\figurenum{1}
\epsscale{0.9}
\plotone{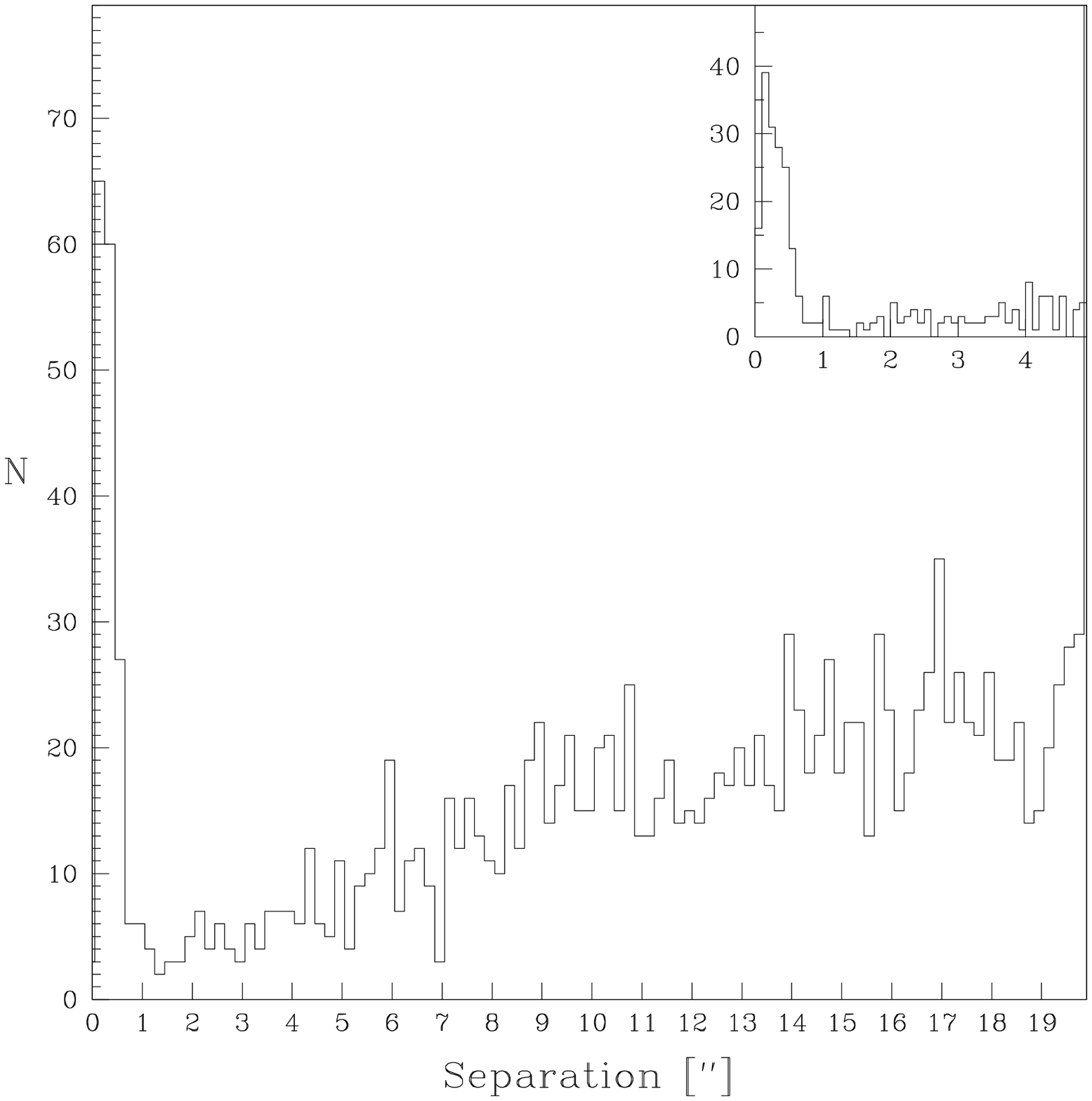}
\caption{Nearest radio source to each optical source in combined SDSS catalogs. Real sources are the peak in center, the randoms are the gradually increasing wing. The inset provides detail for sources with separations less than 5\arcsec.\label{fig-seps}}
\end{figure}

The SDSS photometric database of all detected objects within the A2111 field was correlated with the preliminary radio catalog. The search radius for this correlation was 1\farcs5, as determined by examining the distribution of the separations of each optical source to its nearest radio source (see Figure \ref{fig-seps}). The presence of real associations means this distribution peaks near a separation of zero, then declines rapidly. As the search radius increases, random associations become more likely and the distribution increases steadily. The 1\farcs5 search radius covers the real associations while limiting the number of chance superpositions. The number of chance superpositions in the final list of radio sources with corresponding optical detections was evaluated by determining the average density of objects from the optical catalog around the positions of sources in the radio catalog (note that this is the more appropriate pairing to use, as the surface density of radio sources is a function of distance from the pointing center of the radio observations due to the power pattern response of the VLA). This produced a probability of chance superposition of about 1.7\% for objects separated by 1\farcs5, dropping to less than 1\% for sources separated by under 1\arcsec. 

All sources in the catalog were then analyzed to remove objects with low significance of radio detection. After correcting for the power pattern response of the VLA, the local noise around each radio source was evaluated over a 1\farcm5 box centered on the radio position. The source itself was fit using the task JMFIT, which fits Gaussians to the radio emission and returns the best fit parameters. JMFIT was instructed to correct for the power pattern and also for bandwidth smearing. Sources for which the fitted major axis had a minimum size of zero were considered unresolved and their peak flux measurements were adopted as their flux \citep[e.g.,][]{owen2005}, otherwise the fitted integral flux was adopted. The source flux was then compared to the local noise figure, and sources with fluxes less than five times the local rms were removed. In practice, this removed 16 low significance objects from the list of radio sources with optical IDs.

\begin{figure}
\figurenum{2}
\epsscale{0.9}
\plotone{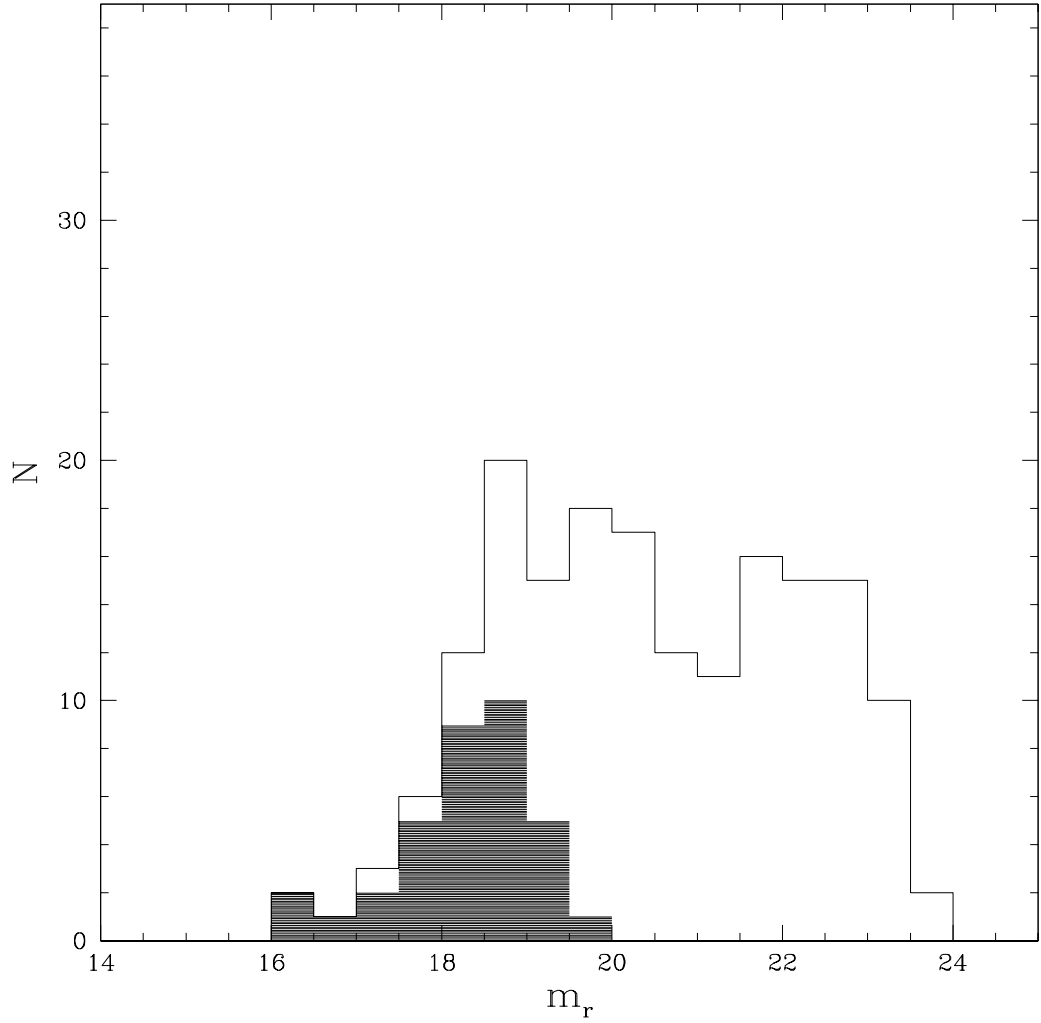}
\caption{Magnitude distribution for radio sources with optical counterparts. The shaded portion represents sources with spectroscopic redshifts.\label{fig-rmags}}
\end{figure}

The list of IDs was then checked by a visual inspection of the actual images. Radio contours were overlaid on the optical images, and all catalog identifications were confirmed. This step also found one radio galaxy with the morphology of a compact double. The separation between the optical galaxy position and the nearer of the two radio lobes was more than 1\farcs5, and hence would be excluded in our radio galaxy catalog. However, using the midpoint of the radio lobe positions results in the source being accepted. This radio galaxy will be discussed in more detail in Section \ref{sec-nonclusterrad}. The final list is presented in Table \ref{tbl-radio} and includes 175 objects. The majority of these were classified as galaxies in the SDSS photometric catalogs, although including objects classified as stars yielded nine detections. Of the optical IDs, 29 are included among our spectroscopic database and a further six had reported spectroscopy from other sources (including one object with an SDSS photometric classification of ``star,'' for which the spectroscopy reveals to be a quasar). This information is included in Table \ref{tbl-radio}, along with photometric redshifts for sources without optical spectroscopy (details on the photometric redshifts will be provided in Section \ref{sec-radgals}). A histogram of the distribution of magnitudes for the radio sources with optical counterparts is provided in Figure \ref{fig-rmags}.

\section{Analysis}

\subsection{Cluster Dynamics from Optical Spectroscopy}\label{sec-dyn}

\begin{figure}
\figurenum{3}
\epsscale{0.9}
\plotone{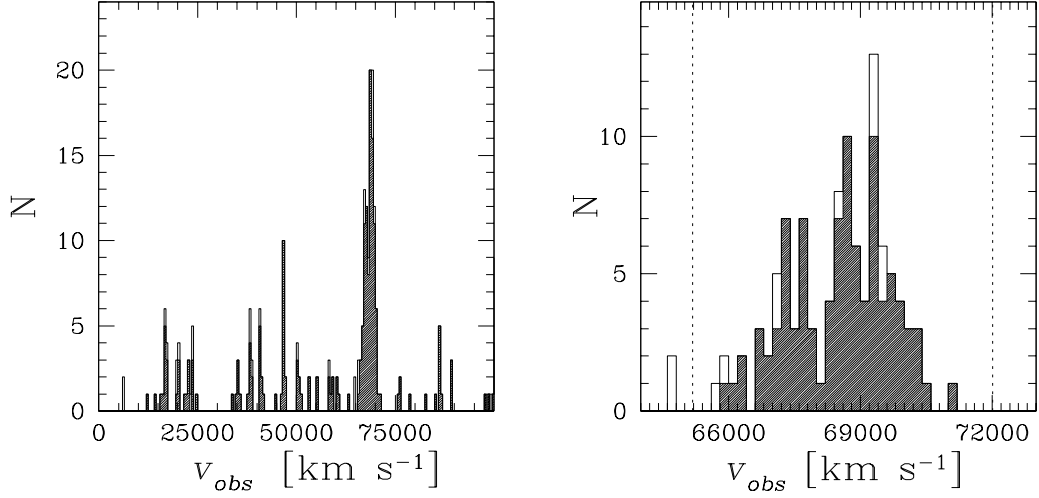}
\caption{Histogram for all galaxies with observed velocity under 100,000 \kms{} (left), and zoomed in to show the cluster (right). Bin sizes are 500 \kms{} for the full histogram and 200 \kms{} for the cluster. The shaded portion represents velocities presented in this paper, with the unshaded portion indicating additional velocities from public sources. Dotted lines indicate the range of accepted cluster velocities on the basis of our velocity measurements.\label{fig-vhist}}
\end{figure}

There are 196 galaxy velocities presented in Table \ref{tbl-vels}, plus an additional 9 targets whose spectra suggest they are faint stars. Velocity histograms along the line of sight to A2111 are provided in Figure \ref{fig-vhist}. Using the robust statistics of \citet{beers1990}, our spectroscopic data indicate that the observed systemic velocity of the cluster (the biweight location) is $68573\pm91$ \kms{} ($z=0.2287$) with a velocity dispersion (the biweight scale) of $889^{+73}_{-59}$ \kms. These values are based on 95 cluster galaxies as determined using 3$\sigma$ clipping, with the galaxy velocities corrected for measurement errors and relativistic effects. Based on these figures, the $\pm3\sigma$ range of observed velocities corresponding to the cluster is 65177 -- 72014 \kms. The biweight location for the cluster center is 15$^{\mbox{{\scriptsize h}}}$39$^{\mbox{{\scriptsize m}}}$42\fs0 +34\degr26\arcmin27\arcsec, about 80\arcsec{} from the adopted X-ray center. Using these data, we derived a virial mass of $3.6 \times 10^{15}$ M$_\odot$. An additional 10 galaxies with velocities from public sources fall within the range of accepted cluster velocities (five each from NED and the SDSS), with a further two galaxies just below the lower velocity limit (see Figure \ref{fig-vhist}). Six of the seven NED velocities, including the two just below the lower velocity limit, come from \citet{lavery1994} and hence may reflect a fairly large offset from our measurements. However, including all additional NED and SDSS velocities does not significantly change the cluster systemic velocity and dispersion, yielding $68536\pm94$ \kms{} and $974^{+75}_{-61}$ \kms, respectively. The two lower velocity galaxies are accepted as cluster members in this extended database on account of their high assigned errors of 300 \kms.

The velocity data were subjected to a wide variety of statistical tests designed to evaluate substructure, as described in \citet{pinkney1996} and references therein. The majority of the tests indicated no significant substructure, although the bimodality hinted at in the velocity histogram (Figure \ref{fig-vhist}) did produce some marginal differences from the null hypothesis of a Gaussian velocity distribution. Substructure tests including positional information failed to indicate significant substructure with the exception of the Fourier Elongation (FE) test \citep{pinkney1996}. This test compares the galaxy distribution to a circularly symmetric distribution, and its significance is evaluated via Monte Carlo simulations. None of the 1000 Monte Carlo runs produced a higher FE test statistic. The elongation is not proof of substructure, but clusters with substructure are typically elongated. In particular, elongation observed in the X-ray emission of the intracluster gas is often viewed as evidence for non-relaxed systems as was noted earlier in the description of prior X-ray results for A2111. The position angle of the elongation in the galaxy distribution, 150\degr$\pm8$ measured east from north, is consistent with that found in the X-ray analysis of \citet{wang1997}, 162\degr$^{+12}_{-21}$ (90\% confidence range).

To further investigate the possible bimodality in the cluster population, we used the KMM algorithm \citep{ashman1994}. The first application of KMM was made on the velocity data only, using the 95 cluster velocities reported in this paper. The input velocities were cosmologically-corrected to the cluster rest frame, and KMM was allowed to fit two Gaussians with identical dispersions to the data (i.e., the homoskedastic case). This fit did largely confirm the impression of bimodality obtained from Figure \ref{fig-vhist}, producing systems with mean velocities of 69177 \kms{} and 67358 \kms, respectively, after conversion back to observed velocities. The fitted dispersion for these two groups was 543 \kms, and 63 galaxies were assigned to the former while 32 were assigned to the latter. The significance of the improvement of this two-component fit over that of a single Gaussian was fairly modest, at about 96\%. The separation of the two components was quite clean and is easily apparent in Figure \ref{fig-vhist}: the trough in the velocity histogram at 68000 \kms{} is exactly where the split occurred, with higher velocity objects belonging to the larger system and lower velocity objects belonging to the smaller one.

\begin{figure}
\figurenum{4}
\epsscale{0.9}
\plotone{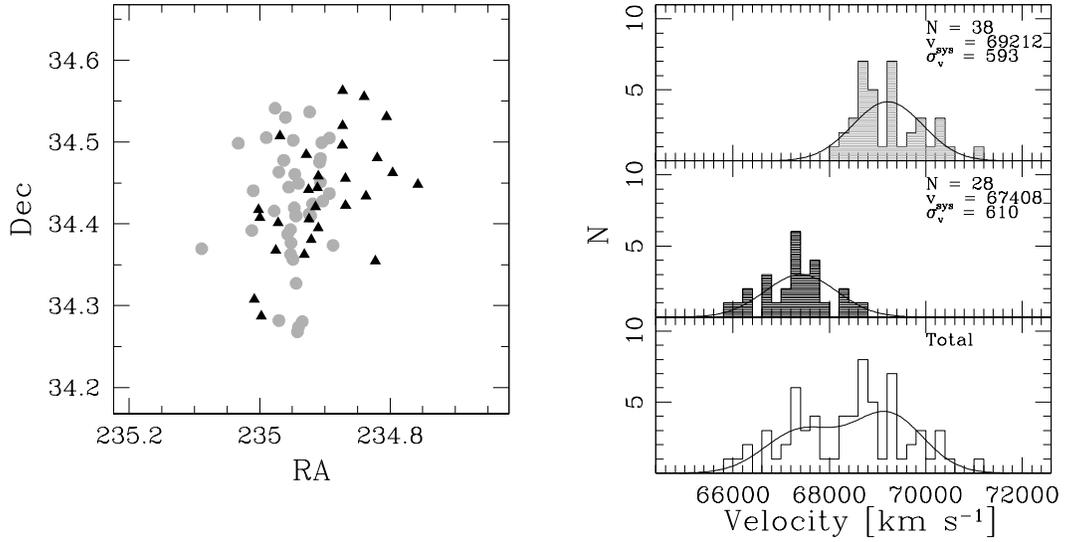}
\caption{Positions of cluster galaxies for the two components fit by KMM (left), and their velocity histograms (right). The grey-shaded circles correspond to the top velocity histogram (also shaded grey), while the black-shaded triangles correspond to the middle histogram (also shaded black).\label{fig-kmm}}
\end{figure}

We also attempted to include positional information in the KMM fitting. In this case, KMM evaluates the probability for each galaxy to belong to either of two components by considering how each variable (two positional coordinates and velocity) compares to the means of those components. The three variables are properly weighted by their individual dispersions and covariances among the variables are taken into account. Provided with the 95 cluster galaxy velocities and their positions, KMM apportioned the galaxies very differently than it did for the homoskedastic case. The two fitted components had similar mean values for their positions and velocities but with very different dispersions in these quantities. Plotting up the positions of the galaxies by fitted component revealed that KMM had taken the core of the cluster and separated it from its outskirts: the core had smaller dispersions in the two positional coordinates and a larger dispersion in velocity. The uncertainty in this split was reflected in a fairly low estimate of correct allocation of galaxies to the ``outskirt'' component (73\%). KMM is known to be sensitive to outliers, so we next restricted the input list to galaxies within 2 Mpc of the cluster center. This resulted in a very similar situation to the homoskedastic case, with two components slightly offset in position (15$^{\mbox{{\scriptsize h}}}$39$^{\mbox{{\scriptsize m}}}$48$^{\mbox{{\scriptsize s}}}$ +34\degr25\arcmin19\arcsec{} and 15$^{\mbox{{\scriptsize h}}}$39$^{\mbox{{\scriptsize m}}}$36$^{\mbox{{\scriptsize s}}}$ +34\degr26\arcmin11\arcsec, each to within an accuracy of about an arcminute) and distinguished more by their large separation in velocity (69212 \kms{} and 67408 \kms{} observed velocities, or about 1400 \kms{} in the cluster rest frame). The fitted dispersions were 593 \kms{} and 610 \kms, with 38 galaxies assigned to the first component and 28 to the second. The actual assignments were very similar to those made in the homoskedastic case: only four galaxies had different assignments in the 3D case from those made in the 1D homoskedastic KMM run. All of these had been assigned to the first component in the homoskedastic fitting but were placed in the second component when positional data were included. Velocity histograms and galaxy positions for these assignments may be found in Figure \ref{fig-kmm}. The two brightest galaxies at the cluster center both belong to the first component. Treating these two components separately, we determined virial masses of $8.5 \times 10^{14}$ and $1.1 \times 10^{15}$ M$_\odot$, respectively.

\subsection{Cluster Blue Fraction}\label{sec-bluefrac}

The availability of accurate photometric data from the SDSS makes computation of the cluster blue fraction a straightforward proposition. In addition, the large spectroscopic database allows for direct testing of whether the high reported blue fraction for A2111 is significantly biased by contamination from foreground and background galaxies, as one might naively expect based on the rich line of sight evidenced in the velocity histogram of Figure \ref{fig-vhist}.

\begin{figure}
\figurenum{5}
\epsscale{0.9}
\plotone{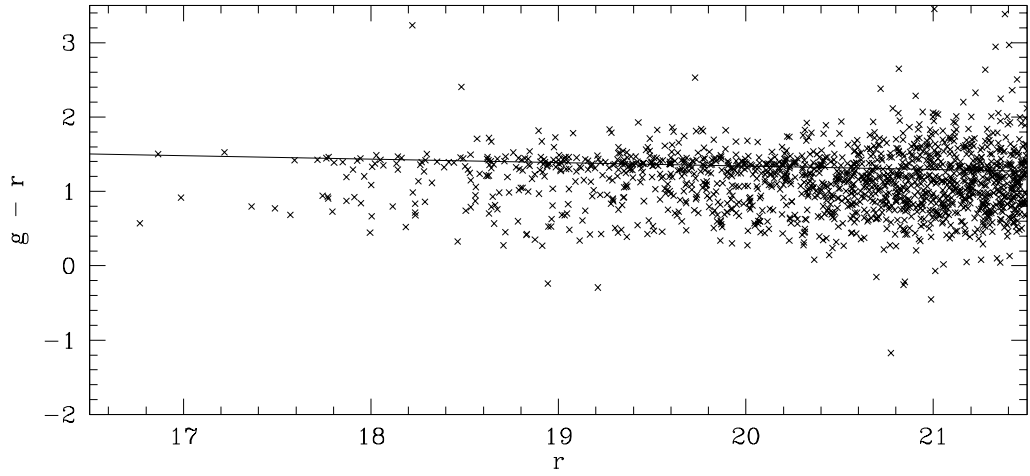}
\caption{Color-magnitude diagram for galaxies in the A2111 field, with the solid line representing the fit of $(g-r) = 2.28 - 0.047r$.\label{fig-cmdfit}}
\end{figure}

\begin{figure}
\figurenum{6}
\epsscale{0.9}
\plotone{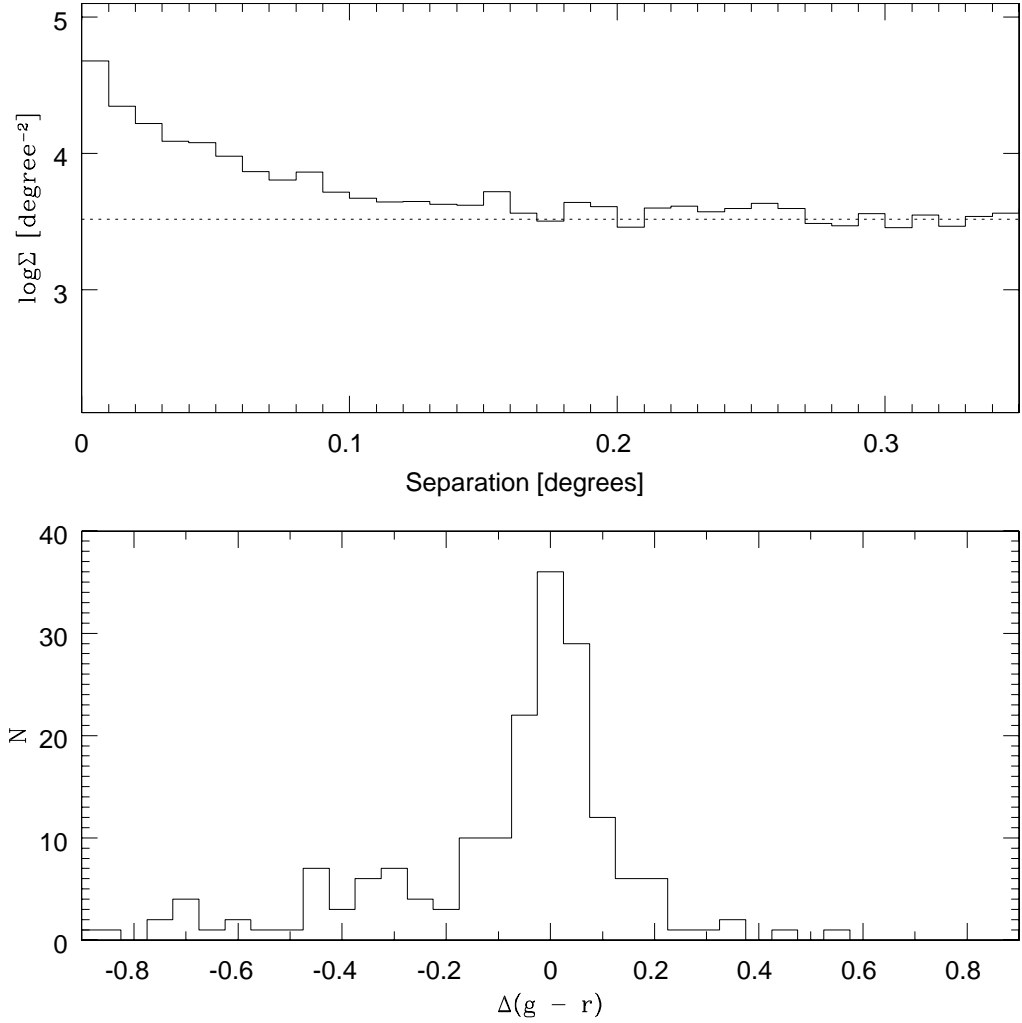}
\caption{Radial distribution of galaxies in the A2111 field brighter than $m_r = 20.68$, with adopted background density of 3290 galaxies degree$^{-2}$ plotted as dotted line (top). $(g-r)$ color relative to the fitted CM relation for such galaxies within 4\farcm1 of the cluster center (bottom). Objects with corrected colors less than -0.2 are considered ``blue.''\label{fig-cdist}}
\end{figure}

We have emulated the original procedure of \citet{butc1984} in evaluating the blue fraction for A2111. In their calculation, BO84 used galaxy counts with $M_V \leq -20$ and a cosmology with $H_0 = 50$ \kms{} Mpc$^{-1}$ and $q_0 = 0.1$. We have used {\ttfamily kcorrect v3.2} \citep{blanton2003} to determine the appropriate k correction for an $M_V = -20$ E/S0 galaxy at $z=0.2287$. The {\ttfamily kcorrect} software also has the feature of synthesizing $B~V~R$ magnitudes from the SDSS $u~g~r~i~z$ magnitudes, making this particularly useful for the purpose at hand. Including $A_V = 0.087$ \citep{schlegel1998}, we determined that $m_r = 20.68$ corresponds to the cutoff of $M_V = -20$. The BO84 definition of blue was a $B - V$ color more than 0.2 magnitudes bluer than the E/S0 galaxies in the cluster. For the SDSS data, this corresponds to 0.2 magnitudes in $g - r$. The next important step is that of fitting the color-magnitude (CM) relation in A2111, also referred to as the E/S0 ridgeline. For this purpose, we have followed the procedure outlined in \citet{lopezcruz2004} and arrived at $(g - r) = 2.28 - 0.047r$ (see Figures \ref{fig-cmdfit} and \ref{fig-cdist}). 

With these results in place, the blue fraction was determined using the SDSS photometric catalog of galaxies. The BO84 blue fraction was calculated inside a radius of 4\farcm1, and we found 183 galaxies with $m_r \leq 20.68$ within this distance of our adopted cluster center position. These 183 galaxies included 45 which were 0.2 magnitudes or more bluer than the fitted CM relation. The background was fit using galaxies at large radial separations from the cluster center where the galaxy surface density had fallen to a constant value. This produced final numbers of 20.1 blue counts and 134.7 total counts, or $f_b = 0.15 \pm 0.03$ (see Figure \ref{fig-cdist}). This is in excellent agreement with the original BO84 estimate of $f_b = 0.16 \pm 0.03$. The lower number of total counts relative to the BO84 value (134.7 compared to 155) was also noted by \citet{lavery1986}, who opted to reduce their magnitude limit to reproduce the net counts of BO84.

The spectroscopic observations confirm this high blue fraction in A2111. There are available redshifts for 44 galaxies in the above analysis, 34 of which are for ``red'' galaxies and 10 of which are for ``blue'' galaxies. The majority of each prove to be cluster members (33 of 34 red galaxies, 9 of 10 blue). Using these fractions to correct the total counts for foreground and background contamination yields $f_b = 0.23 \pm 0.03$.

We have similarly calculated blue fractions for galaxies with $M_r \leq -20$ under the WMAP cosmology within a radius of 2 Mpc (9\farcm1). In this case, there were 466 net galaxy counts of which 156 were bluer than the cluster CM relation. The resulting fraction using backgrounds estimated from the photometry data was $f_b = 0.17 \pm 0.03$. The spectroscopy data include 64 red galaxies within this area, 56 of which are cluster members.\footnote{Since the need for the spectroscopy is simply to indicate cluster membership, we include NED and SDSS velocities in this analysis. This explains the apparent discrepancy in numbers of cluster galaxies within 2 Mpc used for this analysis with that done for the cluster dynamics.} For the blue galaxies, these numbers are 42 and 18, respectively. The resulting blue fraction using spectroscopic data to correct for the background is $f_b = 0.20 \pm 0.03$.

%Note to self: fraction errors assume no error in background figures, just simple poisson on corrected counts. Not sure I like this, but based on figures in papers that is how people do it in the photometry case.

\subsection{Spectroscopic Classifications}

The galaxies with optical spectra were also classified using the ``MORPHS'' scheme \citep{dressler1999,poggianti1999}. This is based primarily on two spectral features, the \oii{} $\lambda3727$ line in emission and the \hd{} $\lambda4102$ line in absorption. Passive galaxies, defined as those lacking emission lines, are designated ``k'' and are further classified on the basis of their \hd{} absorption lines: those with weak \hd{} are simply ``k'' galaxies, while strong \hd{} absorption caused by a large population of young A stars produces a ``k+a'' classification (or if especially strong, ``a+k''). These k+a and a+k galaxies are often referred to as post-starburst or ``E+A'' galaxies \citep{dressler1983}. The presence of \oii{} in emission indicates an active galaxy, with such galaxies being designated ``e()'' where the parenthetical value takes one of four possible options. If examination of line widths and other lines such as \oiii{} indicate an AGN, the galaxy is classified as ``e(n).'' Galaxies with spectra representative of continuous star formation histories evidenced by moderate \oii{} emission and at most slight \hd{} absorption are classified ``e(c),'' while those with unusually strong \oii{} emission representative of a current starburst are classified ``e(b).'' Finally, star forming galaxies with strong \hd{} absorption are classified ``e(a).'' This type of spectrum appears to arise in dusty starburst galaxies \citep{poggianti1999}. 

\begin{figure}
\figurenum{7}
\epsscale{0.9}
\plotone{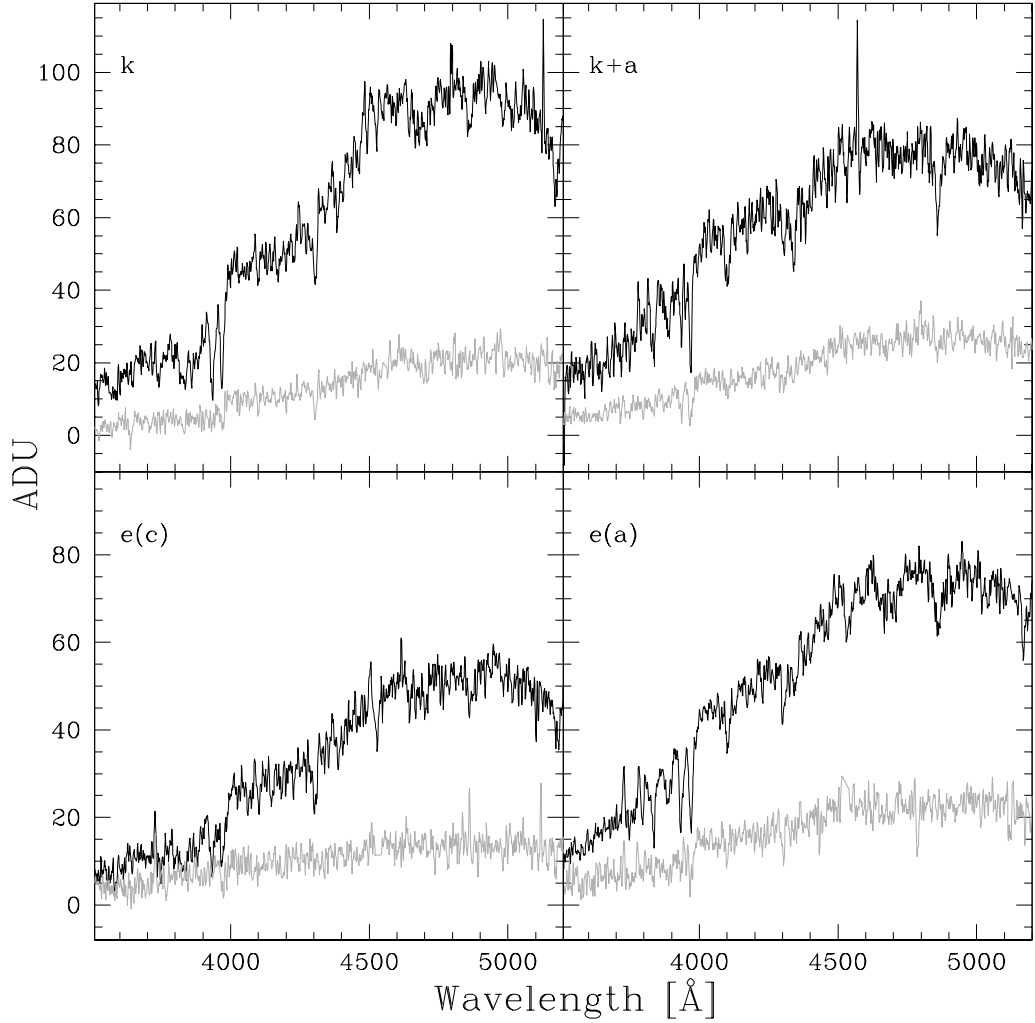}
\caption{Rest-frame spectra for each of the main MORPHS classes of cluster galaxies. The black spectrum for each class represents the highest S/N spectrum while the grey spectrum is the lowest. No flux calibration was performed on the spectra.\label{fig-spec}}
\end{figure}

Our classifications were made within IRAF through visual inspection of the spectra and fitting Gaussians to the appropriate lines. This does imply a certain degree of subjectivity, as the S/N of the spectra are only moderate in the vicinity of \hd{} (a mean of about 12 per resolution element, with a range from 8 to 18) and fitted equivalent widths are influenced by our assumed continuum levels. However, the objective of making the assigments is to assess trends in the star formation history of the cluster galaxies as an overall population. Thus, such issues should have minimal effects. Among our 95 cluster galaxies with Hydra spectra, we identified 63 galaxies of type k, 5 k+a, 14 e(a), 12 e(c), and 1 e(n). The classifications are presented in Table \ref{tbl-spec} and example spectra are presented in Figure \ref{fig-spec}, with both the lowest and highest S/N spectrum within the major classes plotted. The availability of the SDSS spectroscopy allows classification of another five cluster galaxies, all of which are of type k. That all the galaxies with SDSS spectra are type k is to be expected, as the faint limit of the SDSS main galaxy sample is $m_r = 17.77$ \citep{strauss2002} which corresponds to $M_r = -22.9$ for an elliptical in A2111. Some ellipticals fainter than this apparent magnitude limit have SDSS spectra through their selection in the SDSS luminous red galaxy (LRG) sample \citep{eisenstein2001}.

\subsection{Characterizing the Radio Galaxies}

\subsubsection{Cluster Radio Galaxies}\label{sec-radgals}

Table \ref{tbl-props} includes the derived quantities for the 14 radio galaxies whose velocities place them within A2111. The majority of these (9) have optical spectra which indicate their radio emission arises from active star formation. The remaining five radio galaxies include three passive or k type galaxies, one emission line AGN, and one galaxy whose velocity was reported in \citet{morrison2003}. This last galaxy is presumably of type k on the basis of its optical color. The AGN is among the fainter cluster galaxies with an obtained spectrum ($M_R=-21.3$) and is located in a tight grouping of galaxies nearly 2 Mpc to the south of the cluster center.

Radio luminosities for the cluster galaxies were determined using the WMAP cosmology and assuming a spectral index of 0.7 in performing the k correction (i.e., $S_\nu \propto \nu^{-0.7}$). There are no cluster galaxies with extended radio morphologies and luminosities representative of \citet{fanaroff1974} class sources. Star formation rates for the active galaxies were derived using the relationship in \citet{yun2001}, which is based on the local far-infrared (FIR) density and the strong correlation between radio and FIR emission in star-forming galaxies. Underlying assumptions in the FIR-derived SFR, and hence the radio one, are a Salpeter initial mass function with mass limits of 0.1 and 100 M$_\odot$. With this conversion, the sensitivity limit of the radio observations translates to about 5.5 M$_\odot$ year$^{-1}$ (8.5 M$_\odot$ year$^{-1}$ in the noisier regions within 2 Mpc of the cluster center). Thus, the radio observations select the more vigorously star-forming galaxies in the cluster and hence we detect only 6 of the 15 spectroscopically-identified e(a) galaxies and 3 of the 12 e(c) galaxies. Additionally, many of the non-detected emission line galaxies lie farther from the cluster center where the sensitivity of the radio observations are lower. Of the 9 undetected e(a) galaxies, 5 lie over 2 Mpc from the cluster center including one which is outside the region covered by the radio images. Similarly, 7 of the 9 undetected e(c) galaxies are greater than 2 Mpc from the cluster center.

An additional use of the {\ttfamily kcorrect} software is that it can be used to determine photometric redshifts ($z_{phot}$). These are especially useful for determining the properties of the radio galaxies without available spectra, and are presented for such cases in Table \ref{tbl-radio}. The 1$\sigma$ error in these values, determined by comparison of $z_{phot}$ calculated for all galaxies with spectroscopic redshifts presented in Table \ref{tbl-vels}, is 0.044 (after 5 sigma clipping and the $1+z$ correction). From this information we are also able to identify those radio galaxies without optical spectra which are probable cluster members. There are 25 possible cluster radio galaxies within 2 Mpc of the cluster center, of which 8 are photometrically classified as ``blue'' with the remaining 17 being ``red.'' Allowing a $\pm3\sigma$ range, galaxies with $0.10 \leq z_{phot} \leq 0.36$ are potential cluster members and we find that 7 of the 8 blue galaxies meet this criterion while 5 of the 17 red galaxies do. These 12 potential cluster radio galaxies are presented in Table \ref{tbl-pzprops} along with their absolute quantities calculated under the assumption of cluster membership. Most are fairly faint optically, providing a simple explanation for their exclusion in our spectroscopy observations.

\subsubsection{Foreground and Background Radio Galaxies}\label{sec-nonclusterrad}

Radio galaxies not located within the cluster are useful in other investigations. The more powerful radio galaxies, with $L_{1.4} \gtrsim 10^{23}$ W Hz$^{-1}$, are generally found in cluster or group environments. This is particularly true for sources with FR1 morphologies \citep{fanaroff1974}, which have twin jets of radio emission which fade with distance from the host galaxy (``edge darkened''). The bright radio nature of these types of radio galaxies makes them efficient signposts to the presence of galaxy clusters and groups out to high redshifts \citep[e.g.,][]{blanton2001,venemans2002}.

There are 21 radio galaxies with optical spectra indicating they lie in the foreground or background. The properties of these galaxies are provided in Table \ref{tbl-bgprops}. Distances, and hence absolute quantities, are based directly on the measured redshifts; possible peculiar motions of galaxies in clusters are unaccounted. Based on their optical spectra and radio luminosities, most of the galaxies appear to be powered by star formation. The two highest redshift sources, with $z=0.5512$ and $z=0.8718$, are quasars with spectra taken from the SDSS DR4. The MORPHS classifications for foreground and background galaxies are often more uncertain on account of the wavelength coverage of the spectra having been optimized for cluster galaxies. Nevertheless, we provide MORPHS classes for the non-cluster radio galaxies in Table \ref{tbl-bgprops} with flags noting where classes are uncertain due to wavelength coverage or signal-to-noise issues.

\begin{figure}
\figurenum{8}
\epsscale{0.8}
\rotatebox{270}{\plotone{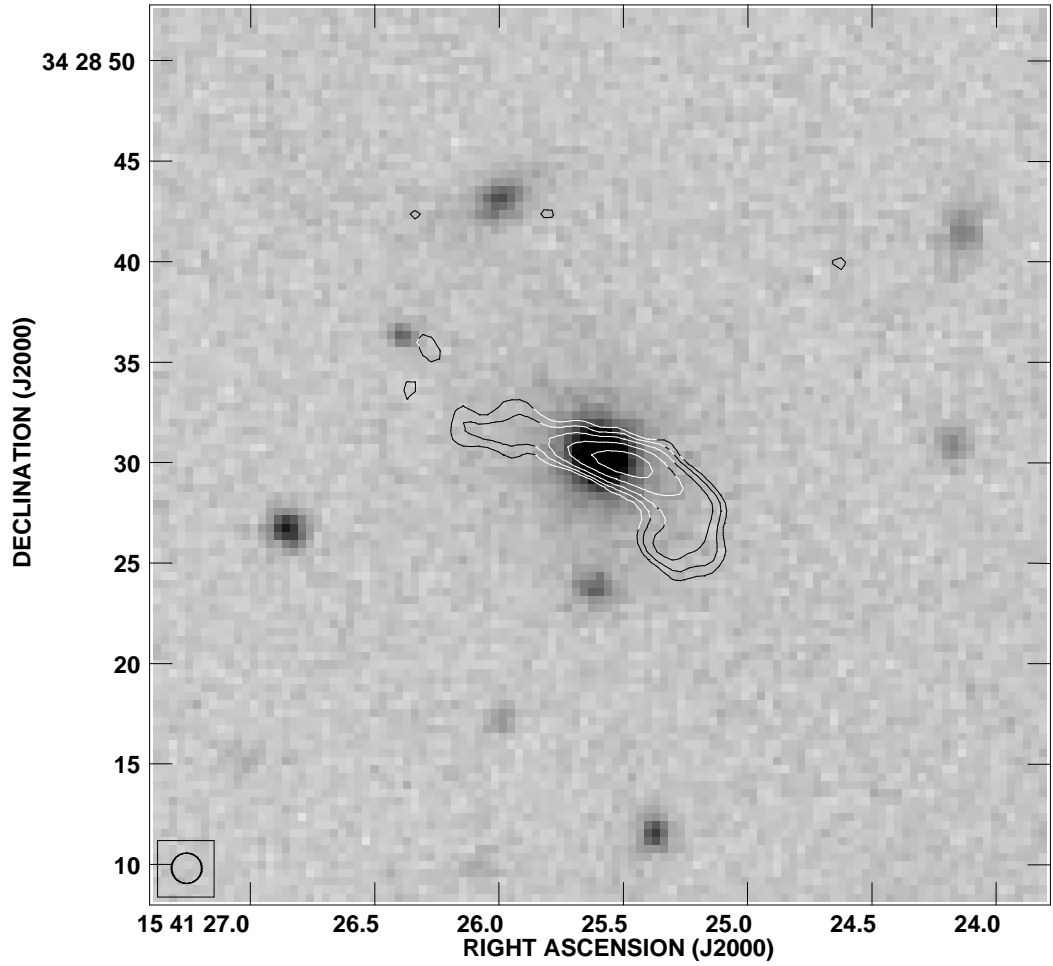}}
\caption{J154125+342830, a foreground radio galaxy with $z=0.1842$. The radio luminosity ($9.1 \times 10^{23}$ W Hz$^{-1}$) and morphology are consistent with the characteristics of typical FR1 sources.\label{fig-J154125}}
\end{figure}

Two foreground galaxies are associated with powerful radio sources and hence are likely markers for poor groups. The galaxy J154125+342830 has the twin jet radio morphology of an FR1 and is presented in Figure \ref{fig-J154125}. Its redshift is $z=0.1842$, producing a radio luminosity of $9.1 \times 10^{23}$ W Hz$^{-1}$ (again assuming $\alpha=0.7$ for the k correction). Its color and optical spectrum are typical of a massive elliptical galaxy as would be expected based on its radio properties. Similarly, J154055+343015 at $z=0.1942$ has a radio luminosity of $1.6 \times 10^{24}$ W Hz$^{-1}$ and the optical spectrum of an elliptical galaxy (although with slight measured \nii{} and \sii). Its location 16\farcm1 from the pointing center of the radio observations means it suffers from smearing effects and consequently its radio morphology is not easily described, although clearly is compact. It is at best marginally resolved by FIRST \citep{first}, with a deconvolved major axis of 1\farcs6 and a flux measurement of 15.34 mJy, consistent with the measured flux from our observations. Neither galaxy has any counterparts within 3000 \kms{} and 1 Mpc projected separation in Table \ref{tbl-vels}.

\begin{figure}
\figurenum{9}
\epsscale{0.8}
\rotatebox{270}{\plotone{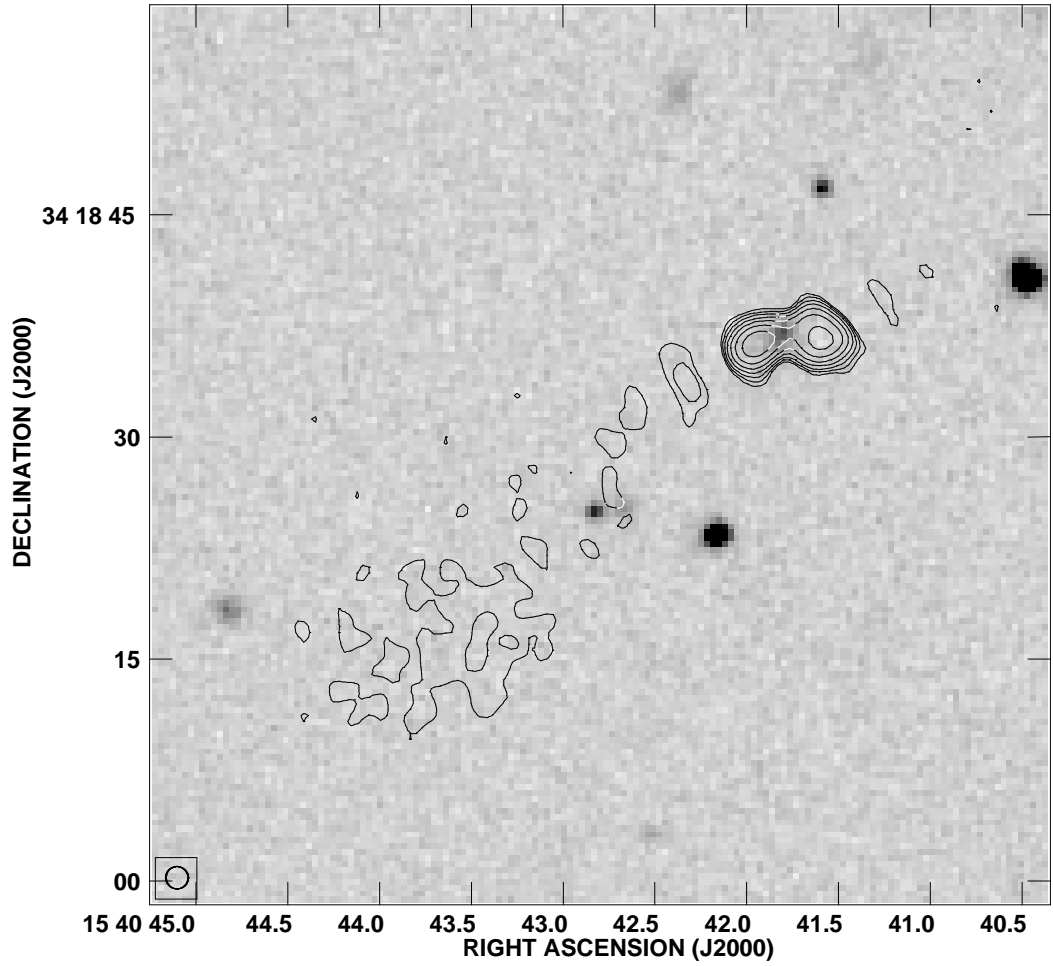}}
%\plotone{J154041.PS}
\caption{J154041+341837, a powerful radio galaxy with a photometric redshift of 0.50. Assuming this redshift, the luminosity of the compact double portion of its emission is $8.7 \times 10^{24}$ W Hz$^{-1}$; inclusion of the diffuse emission to the southeast increases this to $1.7 \times 10^{25}$ W Hz$^{-1}$.\label{fig-J154041}}
\end{figure}

Although it lacks an optical spectrum, J154041+341837 is an interesting extended radio source. This is the compact double noted in Section \ref{sec-ids}, and may be seen in Figure \ref{fig-J154041}. Its photometric redshift is 0.50. In addition to the two main lobes of radio emission, there is diffuse emission to the southeast which appears to connect to the southeast lobe. Including this diffuse emission the system flux is 19.8$\pm$0.5 mJy, equivalent to $1.7 \times 10^{25}$ W Hz$^{-1}$. Without the diffuse emission, the radio flux is still 10.3$\pm$0.2 mJy or a luminosity of $8.7 \times 10^{24}$ W Hz$^{-1}$. These values are consistent with the expectations for radio sources with compact double morphologies at the assumed redshift. 

\section{Discussion}

The objective of our study was to characterize the galaxy population of A2111 and investigate any possible connection it has with the dynamical state of the cluster. We shall turn first to a discussion of the overall cluster dynamics and then investigate the active galaxy populations.

\subsection{Dynamical Assessment of A2111: Evidence for a Cluster Merger?}

The dynamical assessment made from our measured velocities is only suggestive of A2111 representing a cluster merger. Based on our 95 cluster galaxies, few statistical tests indicate significant substructure consistent with a merger scenario. The tests performed exclusively on velocity \citep[i.e., the ``ROSTAT'' package of][]{beers1990} produced a few results indicating a more kurtotic velocity distribution than would be expected should the true velocity distribution be a Gaussian, yet these were only significant at around the 90\% confidence level. Application of the KMM algorithm did indicate that the velocities were better fit by a pair of Gaussians, as might result should A2111 be two separate clusters in the process of merging. However, the significance of the improvement of this fit over a single Gaussian was only a modest 96\%. Inclusion of positional information did little to change this interpretation, as all but two of the 2D and 3D tests described in \citet{pinkney1996} failed to indicate significant substructure at even the 90\% confidence level. For one of these two, the $\Delta$ test \citep{dressler1988}, the marginal significance (98 of 1000 Monte Carlo shuffles of the velocity data while holding the positions fixed produced higher $\Delta$ statistics) is likely not related to substructure so much as sampling. Restricting the analysis to galaxies within 2 Mpc of the cluster center, the signficance declines to about 75\% (255 of 1000 MC runs with higher $\Delta$). The $\Delta$ test uses nearest neighbors to identify local deviations from the overall systemic velocity or dispersion. Much of the higher $\Delta$ for the full sample results from the lower velocity dispersion of galaxies at the periphery of the cluster, which is expected for relaxed systems. The 66 galaxies within 2 Mpc of the cluster center have $\sigma_v = 951^{+96}_{-74}$ \kms, while the 29 outside this radial limit have $\sigma_v = 711^{+118}_{-79}$ \kms.

\begin{figure}
\figurenum{10}
\plotone{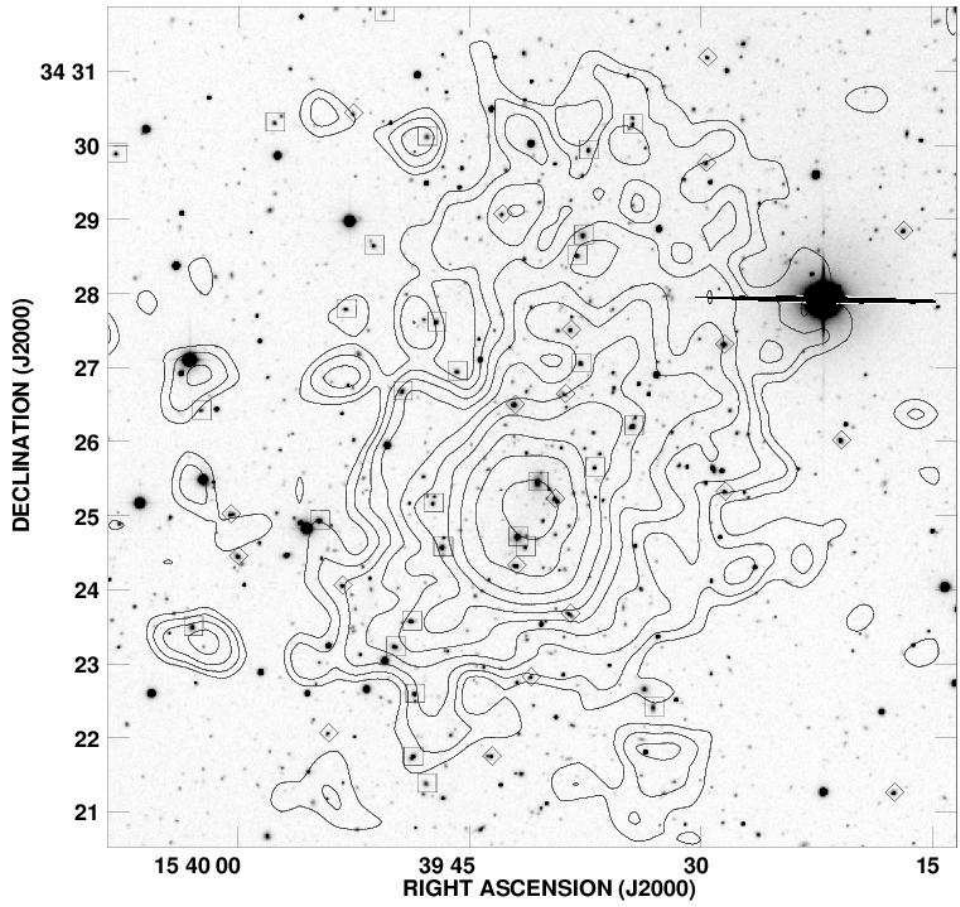}
\caption{$R$-band optical image of cluster center with ROSAT PSPC contours. Galaxies with cluster redshifts are boxed (if first KMM group) or diamonded (if second).\label{fig-xopt}}
\end{figure}

It is interesting, however, that the one significant substructure test matches so well with the X-ray data. The Fourier Elongation (FE) test yielded highly significant evidence for an elongated galaxy distribution, with none of 1000 MC simulations producing a stronger signature of elongation. This result was unchanged when only galaxies within 2 Mpc of the cluster center were considered (in this case, 1 of 1000 MC runs had higher FE). As noted earlier, the derived position angle of the galaxy distribution is consistent with the large-scale X-ray emission. This alignment was noted by \citet{wang1997}, along with its consistency to the alignment of the two brightest galaxies at the cluster core. This overall impression may be seen in Figure \ref{fig-xopt} which depicts the $R$-band optical image with overlaid contours of the X-ray emission from the {\it ROSAT} PSPC observation. Comparison with the galaxy distributions from the KMM analysis (Figure \ref{fig-kmm}) suggests a connection between the extension of the X-ray emission to the northwest of the cluster center and the second fitted group.

Are these results consistent with a head-on cluster merger, the dynamical assessment advocated by \citet{wang1997} and later \citet{henriksen1999}? The merger interpretation of the X-ray data was based on the asymmetric X-ray morphology which leads to shifts in the fitted centroid and ellipticity with scale, along with the presence of a higher temperature core. These results are consistent with hydro/N-body simulations after the subcluster has passed through the core of the primary cluster \citep{roettiger1993}. The simulations of \citet{pinkney1996} indicate that substructure tests performed exclusively on velocities generally do not produce significant results within 2 Gyr of the time when the cores of merging clusters are coincident, provided the difference between the merger axis and the line of sight is greater than about 30\degr. Were the difference between the viewing angle and the merger axis smaller than this amount, the velocity data would almost certainly produce significant deviations from the null hypothesis of a single Gaussian velocity distribution. Folding in the results of the KMM analysis that the two possible merging partners have nearly equal dispersion but are separated in velocity by $\sim1400$ \kms, the substructure tests would argue for the system being viewed shortly after core passage. This same description of the merger timing and geometry can be consistent with the tests which include positional information. The FE statistic is quite good at times near core passage, provided the system is not viewed along the merger axis (i.e., the same constraint resulting from the lack of significant tests based exclusively on velocities). The remaining tests generally do not produce significant results shortly after core passage, with the possible exception of the $\Delta$ test. In summary, while the substructure tests on the galaxy position and velocity data do not themselves provide strong evidence for a cluster merger, they are consistent with the interpretation of the X-ray data that a head-on cluster merger has occurred recently provided that the merger axis is $\gtrsim30\degr$ different from the line of sight. Finally, we note that the merging partners implied by the KMM fit meet the Newtonian condition for a bound system if the difference between the merger axis and the line of sight is between 5\degr{} and 73\degr. Thus, the X-ray and optical data may be explained by a cluster-cluster merger provided the viewing angle is between about 30\degr{} and 70\degr{} from the merger axis and the system is seen near the time when the cores of the progenitor clusters are coincident.

% Seems best gross fit is 60 degrees, 0.2 Gyr post-merger, 3:1 mass ratio. Slight discrepancy with DS test, which would probably be significant for this geometry (but not if the mass ratio were a little higher).

\subsection{Active Galaxies in A2111: Colors, Spectra, and Radio Properties}

The SDSS photometric data and our spectroscopy confirm the high blue fraction for A2111, indentifying it as an active cluster. Similarly, the blue galaxies do appear to be a dynamically young population with a higher velocity dispersion than that of the red galaxy population. The velocity dispersion of the red galaxies from our spectroscopic observations (72 members) is $807^{+77}_{-60}$ \kms{} whereas for the blue galaxies (23 members) it is $1128^{+217}_{-138}$ \kms, a difference significant at just over 2$\sigma$. This general result is unchanged if we consider only those galaxies within 2 Mpc, and suggests that the blue galaxies are more representative of an infalling population which has not virialized. Previous Butcher-Oemler studies have noted the same general effect \citep[e.g.,][and references therein]{lavery1994}. 

The interpretation of the blue galaxies as active galaxies which represent the infalling population does not translate simply to the spectroscopic classifications of the galaxies. In fact, there is no significant difference between the velocity dispersion of the passive, non-emission line galaxies compared to the emission line galaxies. The former have a velocity dispersion of $860^{+85}_{-66}$ \kms{} while the latter have a velocity dispersion of $737^{+128}_{-84}$ \kms. In this case, the effect of including only galaxies within 2 Mpc of the cluster center is larger as the emission line galaxies are more likely to be located at greater clustercentric distance where the local velocity dispersion is lower. There are 27 total emission line galaxies, of which 14 lie outside 2 Mpc. Using only galaxies within 2 Mpc, the velocity dispersion of the emission line galaxies rises to $1038^{+298}_{-160}$ \kms, although this is still consistent with the dispersion of the non-emission line galaxies.

Of course, one reason for any differences in the calculated dispersions is that the simple correspondence of a blue galaxy being an emission line galaxy is often false. Of the 27 cluster emission line galaxies, eight have colors consistent with the cluster red sequence. One of these is the AGN J153945.7+341625, while the other seven have spectra of star-forming galaxies. Star-forming galaxies with colors consistent with a cluster's red sequence have been identified previously by several authors \citep{homeier2005,wolf2005}. There are also exceptions on the converse side, with four non-emission line galaxies having blue colors. Two of these are k+a galaxies, J153938.7+342639 and J153941.9+342420, where the blue color is consistent with the recent burst of star formation implied by the A star features in the galaxies' spectra \citep[the strong Balmer absorption and blue color for these particular galaxies are also noted in ][]{lavery1994}. These two galaxies also lie very close to the center of the cluster, with projected clustercentric distances of 360 and 170 kpc. The remaining two passive cluster galaxies with blue colors are also located near the cluster center (about 0.6 and 1.1 Mpc). Although not detected by the radio data, they are potentially analogs to the cluster star-forming galaxies without notable emission lines in the blue portion of their spectra identified in \citet{miller2002}.

The spectral classifications of the galaxies in combination with the radio data do indicate evidence for starburst activity in member galaxies, with the strong suggestion that dust extinction is important. Although our MORPHS classifications of cluster galaxies include no e(b) galaxies, such types of objects are generally rare in clusters. Of the ten clusters studied by \citet{dressler1999}, most have only 0 to 2 such galaxies. Furthermore, galaxies with e(b) spectra are almost always intrinsically fairly faint \citep{poggianti1999} and our spectroscopy is primarily of brighter cluster galaxies (with $M_R\lesssim-21.5$). This is compounded by our radio detection limit, which would require relatively high SFR for such galaxies. Even so, there may be examples of e(b) galaxies among the radio galaxies with photometric redshifts consistent with cluster membership (Table \ref{tbl-pzprops}), as these were typically missed by our spectroscopy on account of their fainter magnitudes. There are, however, 14 e(a) galaxies identified in A2111, including six associated with radio emission. Models and nearby examples of galaxies with such spectra indicate that e(a) galaxies are dusty starbursts, where emission line regions face larger extinction than the more dispersed young star ($\sim$A) population \citep[e.g.,][]{poggianti1999,shioya2001}. The e(a) galaxies in A2111 include the strongest cluster radio galaxy, J153949.3+342640, an Sc galaxy \citep[morphology from][]{fasano2000} in the cluster core with $L_{1.4} = 9.3 \times 10^{22}$ W Hz$^{-1}$. This translates to a SFR of over 55 M$_\odot$ year$^{-1}$. Finally, two of the three radio-detected e(c) cluster galaxies have colors which place them along the cluster red sequence. If these are also dusty star-forming galaxies as suggested by \citet{wolf2005}, their radio emission implies SFRs of 17 and 27 M$_\odot$ year$^{-1}$.

\citet{smail1999} examined the cluster A851 and identified several k+a galaxies with radio emission, concluding that these were likely highly dust-enshrouded starbursts. Such spectra are essentially an extreme manifestation of the e(a) class, in which dust extinction has removed even the \oii{} line. The rms sensitivity of the \citet{smail1999} data corresponds to a 5$\sigma$ limit of 14.2 M$_\odot$ year$^{-1}$, meaning k+a galaxies in A2111 with comparable SFR would be detected. None of the five k+a galaxies in our spectroscopic data are formal radio detections, however we note that one is identified with low significance. The k+a galaxy J153938.7+342639 is resolved and has an integral flux of $99\pm41$ $\mu$Jy and a local noise of 13 $\mu$Jy, however it did not enter our catalog of possible radio sources as its peak flux was only 40 $\mu$Jy. The integral flux of this galaxy corresponds to a SFR of 8.4 M$_\odot$ year$^{-1}$. It is one of the two blue k+a galaxies near the cluster core.

Are such findings on star formation in A2111 unusual relative to other clusters? First, we compare the populations in the various spectroscopic classes with those of the ten clusters with $0.37 \leq z \leq 0.56$ used in the MORPHS study. In general, the fraction of star-forming galaxies is consistent with the overall results of the MORPHS study. We find that 0.13$\pm$0.03 of the A2111 galaxies have e(c) spectra, whereas the fraction for this class in the MORPHS study was 0.14$\pm$0.02. Similarly, A2111 has an e(a) fraction of 0.15$\pm$0.04 compared to the MORPHS result of 0.11$\pm$0.02. The possible excess in the e(a) population is offset by the noted lack of e(b) galaxies in A2111, which make up 0.05$\pm$0.01 of the MORPHS sample. Our A2111 spectroscopy are more discrepant with the MORPHS results when we consider passive galaxies. In particular, we find relatively more k type galaxies ($0.67\pm0.05$ compared to $0.48\pm0.04$) at the expense of k+a galaxies ($0.05\pm0.02$ compared to $0.21\pm0.02$). In part, some of these minor differences may be caused by sampling as our spectroscopic sampling is $\approx0.5$ magnitude less deep than the MORPHS. Thus, we are less likely to observe the fainter galaxies (especially e(b) galaxies) and will hence have a higher fraction of the brighter spectroscopic classes (i.e., the k galaxies). We do, however, sample a larger physical area than the MORPHS. Based on the radial distributions of the various spectroscopic classes, we would expect this to increase the relative numbers of active galaxies as these are less concentrated in the cluster cores. These arguments do not apply to the apparent deficit in k+a galaxies, however. In the MORPHS work, such spectral types are nearly as bright as the k galaxies and only slightly less centrally concentrated. Only one of the seven MORPHS clusters (Cl1447+23) has a k+a fraction which is not significantly (i.e., at 2$\sigma$) greater than that of A2111. In a general sense, with the exception of the k+a galaxies, the relative fractions of the various spectral classes indicate that despite its lower redshift A2111 has active galaxy populations typical of higher redshift clusters.

Second, we can qualitatively compare the radio galaxy population of A2111 with other clusters studied by radio observations. If we restrict the comparison to galaxies within 2 Mpc of cluster centers and with $L_{1.4}\geq1.4\times10^{22}$ W Hz$^{-1}$ (i.e., 5$\sigma$ detections in the noisier region of our data), there are 11 spectroscopically confirmed cluster radio galaxies in A2111 plus an additional 10 with photometric redshifts suggestive of cluster membership. Inclusion of the photometric redshift sources is probably wise, since the clusters to which we draw comparisons are generally studied down to fainter optical limits. Working in the opposite direction is the known correlation that optically faint galaxies are not often associated with radio sources at luminosities above our noted limit \citep[e.g.,][]{miller2002}. In the highly active cluster A2125 at the comparable redshift of $z=0.2465$, there are 43 galaxies within 2 Mpc and with $L_{1.4}\geq1.4\times10^{22}$ W Hz$^{-1}$ \citep{owen2005}. When factoring in the apparently lower richness of A2125 relative to A2111 \citep[$N_{30}$ values of 62 and 155, respectively, from][]{butc1984}, it is clear that A2111 is less active. At lower redshift, the active cluster A2255 ($z=0.0800$) has 15 radio galaxies meeting the above criteria \citep{my2255}. Only five of these are powered by star formation, with four classified as e(c) and one as e(a). Still, this is about twice the radio galaxy population of the comparable richness Coma cluster.

One possible difference of A2111 from other clusters with large active galaxy populations identified is in its lack of powerful radio AGN, in particular the sources with extended radio morphologies and high luminosities. In fact, there are no FR1 or FR2 type sources in A2111, which is fairly unusual for a rich cluster of galaxies. A851 is the MORPHS cluster best-studied in the radio \citep{morrison2005}, and it also has a large population of radio-detected star-forming galaxies without any FR1 or FR2 galaxies. \citet{venturi2000} suggest that cluster mergers may switch off powerful radio galaxies, based on an apparent lack of such objects in A3558. In contrast, the above mentioned A2255 has six: three head-tail radio galaxies and a strong compact double near the cluster core, plus a wide-angle tail and narrow-angle tail each about 1.5 Mpc from the cluster center.

Finally, we have investigated whether there were any significant differences in the galaxy populations between the two components identified by the KMM analysis. Such a difference might result should one component be associated with an active population merging with a more quiescent population describing the main cluster. We tested the blue fractions, emission line fractions, and radio galaxy fractions and found no statistical difference in these quantities between the two fitted components (for both the fits found using only velocity data and those found using positions as well as velocities). Although not statistically significant, we note that the second component in the KMM fit using a radial cutoff of 2 Mpc and both positional and velocity data included both blue k+a galaxies. The five emission line galaxies assigned to this component were all of the e(a) spectral class and were also all blue. The X-ray data provide circumstantial evidence that this second fitted component is associated with the more diffuse trail of X-ray emission to the northwest of the cluster center (Figures \ref{fig-kmm} and \ref{fig-xopt}).

\section{Conclusions}

We have presented an in-depth analysis of the Butcher-Oemler cluster A2111. This analysis was based on optical spectroscopy providing 196 galaxy spectra, plus multicolor optical imaging from the SDSS and a deep VLA radio continuum image. This combination has been extremely effective in providing a coherent picture of the situation in this cluster.

The optical spectroscopy included 95 confirmed cluster members that allowed us to make a detailed dynamical analysis of the cluster. A large number of statistical tests of substructure were applied, producing only one significant result: the galaxy distribution in A2111 is elongated, with the position angle consistent with that determined from prior X-ray analysis. The velocity data thereby do not themselves provide significant evidence for A2111 being a cluster merger, however they are consistent with this dynamical assessment provided the system is viewed near the time when the cores of the two progenitor clusters have crossed. In addition, the statistical tests required that the viewing angle is $\gtrsim30\degr$ different from the merger axis.

With the SDSS data and our spectroscopy we have confirmed the high blue fraction for A2111. Based on the photometry data alone and using the same aperture as \citet{butc1984}, we derived a blue fraction of $f_b = 0.15 \pm 0.03$. Redshifts are available for 44 of the galaxies that entered this analysis, which revealed that 9 of 10 blue galaxies and 33 of 34 red galaxies are cluster members. Using these figures to account for the background produced $f_b = 0.23 \pm 0.03$, further strengthening A2111's description as a high blue fraction cluster.

The optical spectroscopy and radio data reveal a substantial population of dusty starbursts within the cluster. The sensitivity of the radio data imply that galaxies with SFR as low as 5.5 M$_\odot$ yr$^{-1}$ are detected, and we identified 9 radio galaxies with spectra indicative of star formation from our spectroscopy database. Most of these have ``e(a)'' type spectra, arguing that they are dusty starbursts. Indeed, the strongest radio source in the cluster is of this variety and has a radio SFR of 55 M$_\odot$ yr$^{-1}$. In addition to the radio galaxies with attendant spectroscopy, we identified 12 galaxies whose photometric redshifts are consistent with their being cluster members. Over half of these had blue colors representative of starbursts. Photometric redshifts were presented for all SDSS objects associated with radio emission, in hope that they will prove useful in other studies.

In total, these results are consistent with the hypothesis that dynamically-active clusters are associated with more active member galaxies than relaxed clusters.

\acknowledgments
The authors thank Glenn Morrison and Frazer Owen for valuable discussions, Jason Pinkney and Keith Ashman for sharing their substructure analysis codes, and Sam Barden for his support with the Hydra observations.

Funding for the Sloan Digital Sky Survey (SDSS) has been provided by the Alfred P. Sloan Foundation, the Participating Institutions, the National Aeronautics and Space Administration, the National Science Foundation, the U.S. Department of Energy, the Japanese Monbukagakusho, and the Max Planck Society. The SDSS Web site is http://www.sdss.org/.

The SDSS is managed by the Astrophysical Research Consortium (ARC) for the Participating Institutions. The Participating Institutions are The University of Chicago, Fermilab, the Institute for Advanced Study, the Japan Participation Group, The Johns Hopkins University, the Korean Scientist Group, Los Alamos National Laboratory, the Max-Planck-Institute for Astronomy (MPIA), the Max-Planck-Institute for Astrophysics (MPA), New Mexico State University, University of Pittsburgh, University of Portsmouth, Princeton University, the United States Naval Observatory, and the University of Washington.

This research has made use of the NASA/IPAC Extragalactic Database (NED) which is operated by the Jet Propulsion Laboratory, California Institute of Technology, under contract with the National Aeronautics and Space Administration.

\clearpage

\clearpage

\begin{deluxetable}{l l r r c}
\tablecolumns{5}
\tablecaption{Velocity Data\label{tbl-vels}}
\tablewidth{0pt}
\tablehead{
\colhead{RA(J2000)} & \colhead{Dec(J2000)} & 
\colhead{$cz$} & \colhead{$\Delta cz$} & \colhead{Lines}
}
\startdata
15 37 53.32 & +34 36 34.2 &  70153 &  51 & (\oii) \\
15 37 55.88 & +34 40 12.3 &     10 &  51 & \\
15 37 57.38 & +34 42 34.9 &  69684 &  52 & (\oii) \\
15 38 14.68 & +34 39 45.3 &  69653 &  49 & \\
15 38 27.74 & +34 24 09.0 &  38023 &  45 & \\
15 38 40.34 & +34 30 33.8 &  34504 &  81 & \hb, \ha \\
15 38 42.05 & +34 30 35.5 &  75670 &  66 & (\oii, \hb) \\
15 38 42.07 & +34 31 13.9 &  58540 &  43 & \oii, \hb, \oiii \\
15 38 42.67 & +34 28 26.8 &  51171 &  54 & \oii, \hb, \oiii \\
15 38 45.05 & +34 32 16.2 &  69349\tablenotemark{a} &  51 & \oii, \hb, \oiii \\
15 38 45.10 & +34 35 04.9 &  38688\tablenotemark{a} &  42 & \hb, \oiii, \ha, \nii \\
15 38 45.17 & +34 46 46.3 &  22798 &  41 & \ha, \nii \\
15 38 48.14 & +34 16 16.8 & 116534 &  43 & \oii, \hb, \oiii \\
15 38 48.22 & +34 24 04.0 &  47365 &  46 & \\
15 38 50.00 & +34 32 36.2 &  69220 &  67 & (\oii) \\
15 38 52.44 & +34 31 43.0 &  40933 &  48 & \\
15 38 54.62 & +34 28 24.6 &  70023 &  57 & (\oii, \hb) \\
15 38 55.10 & +34 24 30.4 &  68523 &  49 & \\
15 38 55.16 & +34 25 14.2 &  40770 &  44 & \hb, \oiii, \ha \\
15 38 55.30 & +34 28 29.1 &  46725 &  59 & (\oii) \\
15 38 55.94 & +34 17 29.4 &  60719 &  44 & \oii, \hb, \oiii \\
15 38 57.96 & +34 32 16.8 &    -54 &  75 & \\
15 38 58.13 & +34 35 07.4 &  89318\tablenotemark{c} &  74 & \\
15 38 58.45 & +34 31 54.5 &  23552 &  41 & \hb, \oiii, \ha, \nii \\
15 38 58.50 & +34 12 01.4 &  22064 &  43 & \hb, \oiii, \ha, \nii, \sii \\
15 39 00.10 & +34 35 07.8 &  46752\tablenotemark{a} &  50 & \oii, \hb, \oiii \\
15 39 00.33 & +34 15 12.2 &  86120 &  48 & \oii, \hb, \oiii \\
15 39 00.73 & +34 33 13.8 &  38725 &  42 & \hb, \oiii, \ha \\
15 39 01.75 & +34 26 54.0 &  68303 &  51 & \\
15 39 02.30 & +34 21 14.8 &  46651 &  48 & (\oii) \\
15 39 03.48 & +34 23 14.3 &  20364 &  41 & \hb, \oiii, \ha \\
15 39 04.77 & +34 19 18.1 &  38193\tablenotemark{a,c}&  41 & \hb, \ha, \nii \\
15 39 05.14 & +34 27 19.6 &  76199 &  53 & \\
15 39 05.82 & +34 31 34.1 &  68739 &  52 & \\
15 39 05.98 & +34 33 06.4 &  67198 &  60 & \\
15 39 07.08 & +34 15 47.5 &  86222 &  76 & \\
15 39 07.54 & +34 35 47.9 &  66962 &  52 & \\
15 39 08.62 & +34 24 57.2 &  44627 &  69 & (\hb, \oiii) \\
15 39 09.07 & +34 25 56.6 &     10 &  57 & \\
15 39 11.06 & +34 27 43.6 &  67434 &  52 & \\
15 39 11.23 & +34 19 10.2 & 100572 &  51 & \oii, \hb, \oiii \\
15 39 11.38 & +34 17 51.7 &  86265 &  46 & \oii, \hb \\
15 39 13.30 & +34 31 50.5 &  68533 &  48 & \\
15 39 14.83 & +34 33 25.2 &  67388 &  51 & \\
15 39 15.58 & +34 30 04.3 &  46642 &  44 & \oii, \hb, \oiii \\
15 39 16.51 & +34 30 09.7 &  46663 &  41 & \oii, \hb, \oiii \\
15 39 16.58 & +34 32 06.4 &  46688\tablenotemark{a} &  44 & \oii, \hb, \oiii \\
15 39 16.78 & +34 28 50.9 &  68737 &  49 & \\
15 39 17.43 & +34 21 15.5 &  68322 &  48 & \\
15 39 19.87 & +34 34 29.4 &  46569 &  42 & \oii, \hb, \oiii \\
15 39 20.07 & +34 14 12.8 &  67948 &  66 & \\
15 39 20.84 & +34 26 01.0 &  66726 &  52 & \\
15 39 21.32 & +34 14 14.3 &  35204\tablenotemark{c} &  52 & \\
15 39 21.50 & +34 33 19.8 &  67175\tablenotemark{a} &  58 & \oii, \hb, \oiii \\
15 39 21.94 & +34 20 00.6 &  86321 &  67 & (\oii) \\
15 39 24.86 & +34 34 44.1 &  68796\tablenotemark{c} &  47 & \\
15 39 27.91 & +34 17 59.3 &  35092 &  47 & \\
15 39 28.20 & +34 31 01.5 &  33960\tablenotemark{a} &  42 & \hb, \ha, \nii \\
15 39 28.38 & +34 25 19.6 &  67942 &  54 & \\
15 39 28.42 & +34 27 19.2 &  67297 &  53 & \\
15 39 29.11 & +34 25 38.6 &  40692\tablenotemark{a,c} &  48 & \hb, \oiii, \ha \\
15 39 29.46 & +34 33 46.4 &  67711 &  54 & \\
15 39 29.47 & +34 31 11.7 &  67521 &  51 & \\
15 39 29.59 & +34 29 46.0 &  66245 &  82 & (\oii, \hb) \\
15 39 30.02 & +34 04 14.9 &  19812 &  49 & \\
15 39 30.26 & +34 50 36.6 &  50277\tablenotemark{a} &  51 & \oii, \hb, \oiii \\
15 39 31.99 & +34 53 50.5 &  69460 &  62 & (\oii) \\
15 39 33.02 & +34 22 25.0 &  68915\tablenotemark{b} &  49 & (\oii) \\
15 39 34.31 & +34 26 12.8 &  68632 &  49 & \\
15 39 34.32 & +34 30 17.4 &  68978 &  59 & \\
15 39 34.33 & +34 30 22.7 &  40604 &  41 & \hb, \oiii, \ha \\
15 39 36.81 & +34 25 39.4 &  69314 &  54 & \\
15 39 37.20 & +34 29 56.4 &  70567 &  51 & \\
15 39 37.58 & +34 28 47.6 &  69893 &  54 & \\
15 39 37.71 & +34 27 03.6 &  68811 &  50 & \\
15 39 37.68 & +34 35 58.9 &  67707 &  41 & \oii, \hb \\
15 39 37.78 & +34 13 24.0 &  38150 &  44 & \hb, \oiii, \ha, \nii \\
15 39 37.93 & +34 28 31.1 &  70291 &  51 & \\
15 39 38.37 & +34 27 30.6 &  67271 &  54 & \\
15 39 38.42 & +34 23 40.2 &  66187 &  62 & (\oii) \\
15 39 38.50 & +34 19 10.6 &  38111\tablenotemark{a} &  41 & \hb, \oiii, \ha, \nii \\
15 39 38.75 & +34 26 38.8 &  67342\tablenotemark{b} &  62 & \\
15 39 39.40 & +34 25 13.8 &  66287 &  54 & \\
15 39 39.89 & +34 21 06.8 &    124 &  70 & \\
15 39 40.46 & +34 19 55.6 &  78784 &  52 & \oii, \hb, \oiii \\
15 39 40.50 & +34 25 27.5 &  68407\tablenotemark{b,c} &  49 & \\
15 39 40.99 & +34 22 49.6 &  67651 &  51 & \\
15 39 41.20 & +34 01 14.2 &  37929\tablenotemark{a} &  45 & \oiii, \ha, \nii, \sii \\
15 39 41.32 & +34 24 34.6 &  68760 &  51 & \\
15 39 41.52 & +34 17 14.6 &  60466 &  61 & \\
15 39 41.67 & +34 32 11.9 &  69146 &  49 & \\
15 39 41.81 & +34 24 43.2 &  68723\tablenotemark{b} &  52 & \\
15 39 41.91 & +34 24 19.8 &  65808\tablenotemark{b} &  60 & \\
15 39 42.01 & +34 26 30.1 &  67704\tablenotemark{c} &  51 & \\
15 39 42.82 & +34 29 04.6 &  67691 &  50 & \\
15 39 43.52 & +34 21 45.0 &  67293 &  53 & \\
15 39 44.21 & +34 20 46.0 &  99807 &  70 & \oii, \hb, \oiii \\
15 39 44.31 & +34 16 49.4 &  70355 &  48 & \\
15 39 45.74 & +34 16 25.3 &  69956\tablenotemark{a} &  48 & \oii, \hb, \oiii \\
15 39 45.74 & +34 26 57.1 &  68718 &  51 & \\
15 39 46.15 & +34 16 05.2 &  68748 &  54 & \\
15 39 46.58 & +34 19 37.6 &  69786 &  55 & \\
15 39 46.68 & +34 24 34.6 &  68166 &  48 & \\
15 39 47.09 & +34 27 37.1 &  71002 &  52 & \\
15 39 47.30 & +34 25 10.2 &  69210 &  59 & \\
15 39 47.70 & +34 30 07.6 &  69236 &  81 & \\
15 39 47.74 & +34 21 23.0 &  69323 &  58 & \\
15 39 48.50 & +34 22 35.8 &  68317 &  48 & \\
15 39 48.60 & +34 21 44.6 &  68761 &  49 & (\oii) \\
15 39 48.71 & +34 23 34.4 &  69225 &  53 & \\
15 39 48.94 & +34 35 37.7 &    -27 &  62 & \\
15 39 49.30 & +34 26 40.9 &  68930\tablenotemark{b} &  66 & (\oii, \hb) \\
15 39 49.77 & +34 23 13.9 &  69428 &  40 & \\
15 39 50.18 & +34 19 57.4 &  85485 &  58 & \oii, \hb, \oiii \\
15 39 50.48 & +34 31 48.4 &  69312 &  56 & (\oii) \\
15 39 51.05 & +34 15 34.2 &  53001 &  50 & \\
15 39 51.15 & +34 28 39.0 &  68252 &  51 & (\oii) \\
15 39 51.26 & +34 36 11.2 &  69055 &  53 & \\
15 39 51.46 & +34 34 41.9 &  98510 &  75 & (\oii) \\
15 39 52.50 & +34 30 26.3 &  67039 &  49 & \oii, \hb, \oiii \\
15 39 52.94 & +34 16 54.5 &  68537 &  49 & \\
15 39 52.96 & +34 27 47.5 &  68868 &  51 & (\oii) \\
15 39 53.19 & +34 24 03.6 &  66886 &  52 & \\
15 39 53.83 & +34 15 47.9 &  46597 &  44 & \oii, \hb, \oiii \\
15 39 54.10 & +34 22 03.7 &  67377 &  62 & \\
15 39 54.34 & +34 32 28.0 &  69959 &  52 & \\
15 39 54.67 & +34 24 56.2 &  68720 &  48 & \\
15 39 54.72 & +34 18 07.2 &  50133 &  49 & \oii, \hb, \oiii \\
15 39 55.08 & +34 21 07.9 &  22839 &  53 & \hb, \oiii, \ha \\
15 39 55.40 & +34 16 54.1 &  86704 &  59 & \\
15 39 55.56 & +34 20 25.4 &  82961 &  59 & \\
15 39 56.54 & +34 15 06.1 &  46667\tablenotemark{c} &  44 & (\oii, \hb) \\
15 39 56.81 & +34 24 28.1 &  55139 &  50 & \\
15 39 57.02 & +34 15 42.1 &  86014 &  64 & \\
15 39 57.59 & +34 30 18.7 &  70101 &  61 & \\
15 39 57.60 & +34 45 46.1 &  41142\tablenotemark{a} &  43 & \hb, \oiii, \ha \\
15 39 58.01 & +34 20 40.6 &  16873 &  51 & \\
15 39 58.20 & +34 46 03.4 &  20004 &  42 & \hb, \oiii, \ha, \nii, \sii \\
15 39 59.14 & +34 17 43.1 &  59393 &  47 & \oii, \hb, \oiii \\
15 39 59.44 & +34 17 12.8 &  67356 &  52 & \\
15 39 59.96 & +34 24 27.4 &  66682 &  54 & \\
15 40 00.48 & +34 25 00.8 &  66649 &  57 & \\
15 40 01.94 & +34 18 26.6 &  67591 &  64 & (\oii) \\
15 40 02.34 & +34 26 25.4 &  68429 &  59 & \\
15 40 02.88 & +34 23 29.8 &  70284\tablenotemark{a} &  43 & \oii, \hb, \oiii \\
15 40 03.12 & +34 34 33.2 &  67755 &  54 & \\
15 40 04.32 & +34 24 59.4 &  97821 &  44 & \oii, \hb, \oiii \\
15 40 05.60 & +34 15 20.2 &  67971 &  51 & \\
15 40 06.03 & +34 33 38.5 &  20024\tablenotemark{a} &  41 & \hb, \oiii, \ha, \nii, \sii \\
15 40 07.87 & +34 29 53.5 &  69294 &  79 & \\
15 40 08.96 & +34 16 03.7 &  69613 &  64 & (\oii) \\
15 40 09.96 & +34 12 53.6 &  14416\tablenotemark{c} &  42 & \hb, \oiii, \ha, \nii, \sii \\
15 40 10.13 & +34 30 53.3 &  60088 &  57 & (\oii) \\
15 40 10.80 & +34 31 16.0 &  59149 &  49 & \\
15 40 10.94 & +34 15 13.3 &  69199 &  54 & (\oii) \\
15 40 13.93 & +34 48 34.9 &  41895 &  48 & \\
15 40 14.13 & +34 18 52.2 &  24899 &  45 & \hb, \oiii, \ha, \nii \\
15 40 14.18 & +34 31 58.4 &  17026\tablenotemark{c} &  43 & \\
15 40 14.45 & +34 30 05.4 &  16926\tablenotemark{a,c} &  49 & \hb, \ha, \nii, \sii \\
15 40 14.93 & +34 27 38.2 &  40661 &  58 & \\
15 40 15.05 & +34 33 37.4 &  16398 &  41 & \hb, \oiii, \ha, \sii \\
15 40 15.10 & +34 34 41.5 &  35813 &  86 & \\
15 40 17.35 & +34 25 48.0 &  47217\tablenotemark{c} &  47 & \\
15 40 17.78 & +34 10 20.3 &  21913\tablenotemark{c} &  49 & \ha, \nii \\
15 40 20.54 & +34 13 12.9 &   -191 &  66 & \\
15 40 21.41 & +34 22 09.8 &  69687 &  48 & \\
15 40 21.60 & +34 35 46.0 &    -26 &  58 & \\
15 40 22.25 & +34 21 31.3 &  23861\tablenotemark{a} &  42 & \hb, \nii, \ha \\
15 40 22.34 & +34 20 42.0 &  76059 &  57 & \\
15 40 23.30 & +34 33 57.2 &  89358 &  79 & \oii, \oiii \\
15 40 23.44 & +34 17 26.5 &  69180 &  59 & \oii, \hb \\
15 40 23.64 & +34 25 22.4 &  50270 &  55 & \\
15 40 24.74 & +34 27 23.0 &  22601\tablenotemark{c,d} &  50 & \ha, \nii, \sii \\
15 40 25.58 & +34 13 21.4 & 100054 &  54 & \oii, \hb, \oiii \\
15 40 25.58 & +34 35 35.5 &  16887\tablenotemark{c} &  44 & \\
15 40 25.94 & +34 25 15.6 &  50619 &  53 & \\
15 40 26.08 & +34 18 40.3 &  46627 &  51 & \oii, \hb, \oiii \\
15 40 26.28 & +34 27 22.0 &  16971 &  56 & \\
15 40 27.19 & +34 16 34.5 &  69567 &  52 & (\oii) \\
15 40 27.94 & +34 13 37.7 &  69552 &  52 & (\oii) \\
15 40 28.08 & +34 17 16.8 &  58388 &  73 & (\oii, \hb) \\
15 40 28.08 & +34 31 34.7 & 114780 &  64 & \oii, \hb \\
15 40 28.16 & +34 29 29.4 &  15655\tablenotemark{a} &  41 & \hb, \oiii, \ha, \nii, \sii \\
15 40 28.54 & +34 15 00.4 &  53010 &  75 & (\oii) \\
15 40 28.97 & +34 33 40.0 &  12024\tablenotemark{a,c} &  41 & \hb, \oiii, \ha, \nii, \sii \\
15 40 29.18 & +34 27 22.0 &  69339 &  57 & (\oii) \\
15 40 29.78 & +34 34 44.4 &     12 &  51 & \\
15 40 30.07 & +34 09 10.9 &  23994 &  44 & \hb, \oiii, \ha, \nii, \sii \\
15 40 30.21 & +34 18 47.5 &  46456 &  48 & \\
15 40 30.92 & +34 30 50.4 &  68474 &  48 & \\
15 40 31.86 & +34 21 01.8 &  35009 &  43 & \hb, \oii, \ha \\
15 40 31.97 & +34 29 19.0 &  67642 &  64 & \\
15 40 32.20 & +34 15 58.8 &  17114\tablenotemark{c} &  68 & (\ha, \nii) \\
15 40 33.25 & +34 33 24.1 &  58464\tablenotemark{a} &  63 & \oii, \hb, \oiii \\
15 40 33.55 & +34 36 18.7 &  50964 &  48 & (\oii) \\
15 40 33.62 & +34 34 53.0 &  89188 &  78 & \\
15 40 35.03 & +34 18 44.6 &    -20 &  53 & \\
15 40 36.07 & +34 04 52.3 &  68526 &  47 & \\
15 40 45.81 & +34 43 57.0 &  63181 &  52 & \\
15 40 52.10 & +34 27 37.8 &  68975\tablenotemark{c} &  48 & \\
15 41 08.98 & +34 27 32.4 &  16967\tablenotemark{c} &  40 & \hb, \ha, \nii, \sii \\
15 41 14.06 & +34 42 46.0 &  41073 &  47 & \\
15 41 25.58 & +34 28 30.4 &  55213 &  47 & \\
15 41 45.24 & +34 21 39.2 &  69835 &  57 & (\oii, \hb)\\
15 41 47.78 & +34 26 24.0 &  17059 &  43 & \hb, \ha, \nii, \sii \\
\enddata

\tablenotetext{a}{A consistent velocity was found for this galaxy via cross correlation of absorption features with velocity standards.}

\tablenotetext{b}{A velocity for this galaxy is also reported in NED.}

\tablenotetext{c}{A velocity for this galaxy is also in the SDSS DR4.}

\tablecomments{All reported velocities are heliocentric. When a velocity has been determined on the basis of emission lines, these lines are noted in the ``Lines'' column. If identified lines are offset by parentheses, they were not used to determine the reported velocity but were found to be consistent with that velocity. Note that \oiii{} usually refers to the $\lambda$5007 line, but sometimes also includes the $\lambda$4959 line. Similarly, \nii{} is usually the $\lambda$6584 line but may also include the $\lambda$6548 line and \sii{} may be either or both of $\lambda$6717 and $\lambda$6731.}

\end{deluxetable}

\begin{deluxetable}{l l l r r r r c r}
\tablecolumns{9}
\tablecaption{Radio Sources with Optical Counterparts\label{tbl-radio}}
\tablewidth{0pt}
\tablehead{
\colhead{RA(J2000)} & \colhead{Dec(J2000)} & \colhead{$z$} & \colhead{$m_r$} & 
\colhead{$S_{1.4}$} & \colhead{$\Delta S_{1.4}$} & \colhead{rms} & \colhead{Res} & 
\colhead{Sep}
}
\startdata
15 38 08.90 & 34 22 29.6 & 0.01   & 21.56\tablenotemark{a} &  330 &  71 & 39 & U & 0.4 \\
15 38 09.55 & 34 35 36.2 & 0.80   & 22.57 & 1046 & 106 & 52 & U & 0.5 \\
15 38 25.22 & 34 34 33.6 & 0.40   & 19.42 &  310 &  61 & 34 & U & 0.5 \\
15 38 27.46 & 34 31 28.6 & 0.86   & 21.82 &  283 &  46 & 28 & U & 0.2 \\
15 38 29.66 & 34 20 22.6 & 0.37   & 22.41 &  594 &  56 & 31 & U & 0.5 \\
15 38 33.31 & 34 30 50.8 & 0.44   & 19.74 &  491 &  38 & 24 & U & 0.4 \\
15 38 34.06 & 34 15 51.8 & 0.43   & 20.65 &  408 & 137 & 37 & R & 0.7 \\
15 38 34.25 & 34 30 32.8 & 0.38   & 22.32 &  230 &  60 & 24 & R & 0.1 \\
15 38 38.88 & 34 18 22.7 & 0.90   & 22.64 &  461 &  49 & 40 & U & 0.5 \\
15 38 39.05 & 34 26 30.1 & 0.80   & 21.69 &  142 &  31 & 21 & U & 0.5 \\
15 38 39.82 & 34 31 08.0 & 0.25   & 19.00 &  219 &  34 & 22 & U & 0.3 \\
15 38 40.53 & 34 23 44.9 & 0.8718\tablenotemark{b} & 18.67\tablenotemark{a} &  184 &  52 & 19 & R & 0.3 \\
15 38 40.94 & 34 33 09.4 & 0.70   & 22.93 &  288 &  37 & 23 & U & 0.1 \\
15 38 43.30 & 34 28 51.2 & 0.46   & 22.60 &  148 &  28 & 21 & U & 0.5 \\
15 38 43.58 & 34 37 13.1 & 0.47   & 18.60 & 1594 &  49 & 29 & U & 0.1 \\
15 38 45.10 & 34 35 04.9 & 0.1292 & 18.41 &  190 &  55 & 26 & R & 0.2 \\
15 38 45.26 & 34 11 54.2 & 0.10   & 18.17 &  356 &  58 & 30 & U & 0.1 \\
15 38 49.25 & 34 17 29.4 & 1.01   & 22.19 &  702 &  90 & 26 & R & 0.3 \\
15 38 49.90 & 34 23 18.2 & 0.63   & 20.76 &  207 &  24 & 18 & U & 0.2 \\
15 38 52.44 & 34 31 43.0 & 0.1365 & 17.87 &  114 &  32 & 22 & U & 0.3 \\
15 38 55.58 & 34 21 43.9 & 0.14   & 22.73\tablenotemark{a} &  795 &  25 & 27 & U & 0.1 \\
15 38 55.94 & 34 26 29.4 & 0.37   & 19.53 &  171 &  40 & 19 & R & 0.1 \\
15 38 57.53 & 34 22 19.6 & 0.92   & 21.05 &  172 &  24 & 21 & U & 0.1 \\
15 39 00.24 & 34 25 47.6 & 0.63   & 21.54 &  393 &  23 & 18 & U & 0.4 \\
15 39 01.06 & 34 15 17.3 & 0.56   & 22.59 &  207 &  31 & 21 & U & 0.3 \\
15 39 01.87 & 34 30 50.0 & 0.68   & 23.05 &  114 &  39 & 19 & U & 0.2 \\
15 39 01.94 & 34 25 46.2 & 0.47   & 20.16 &   94 &  21 & 17 & U & 0.1 \\
15 39 02.33 & 34 16 33.6 & 0.54   & 23.01 &  108 &  26 & 19 & U & 0.5 \\
15 39 03.53 & 34 38 16.1 & 0.49   & 18.29 &  216 &  43 & 26 & U & 0.3 \\
15 39 04.80 & 34 12 26.3 & 0.69   & 22.72 &  249 &  38 & 24 & U & 0.3 \\
15 39 05.83 & 34 28 35.4 & 0.18   & 19.81 &  120 &  22 & 20 & U & 0.3 \\
15 39 06.67 & 34 39 28.8 & 0.1281\tablenotemark{b} & 17.24 &  407 &  84 & 27 & R & 0.4 \\
15 39 07.39 & 34 29 07.4 & 0.68   & 21.15 &  244 &  22 & 23 & U & 0.5 \\
15 39 08.14 & 34 21 10.4 & 0.62   & 20.82 &  701 &  24 & 16 & U & 0.2 \\
15 39 08.62 & 34 24 57.2 & 0.1487 & 18.19 &  220 &  52 & 16 & R & 0.4 \\
15 39 08.98 & 34 29 16.4 & 1.01   & 22.18 &  134 &  21 & 24 & U & 0.3 \\
15 39 10.27 & 34 28 09.8 & 0.84   & 22.13 &  153 &  40 & 19 & R & 0.2 \\
15 39 10.58 & 34 35 37.0 & 0.43   & 20.32 &  167 &  29 & 19 & U & 0.2 \\
15 39 10.61 & 34 38 02.0 & 0.20   & 18.60 &  498 &  37 & 23 & U & 0.2 \\
15 39 11.38 & 34 17 51.7 & 0.2877 & 19.22 &  205 &  38 & 17 & R & 0.1 \\
15 39 12.34 & 34 11 23.3 & 0.51   & 20.84 &  223 &  36 & 24 & U & 0.4 \\
15 39 12.43 & 34 38 26.2 & 0.00   & 21.60 & 1312 &  36 & 27 & U & 0.2 \\
15 39 16.51 & 34 30 09.7 & 0.1557 & 18.96 &  257 &  43 & 22 & R & 0.4 \\
15 39 16.78 & 34 28 50.9 & 0.2293 & 18.68 &  152 &  37 & 21 & R & 0.3 \\
15 39 17.09 & 34 39 05.8 & 0.20   & 18.70 &  184 &  37 & 25 & U & 0.6 \\
15 39 20.28 & 34 32 24.4 & 0.56   & 22.36 &  216 &  20 & 16 & U & 0.3 \\
15 39 21.50 & 34 33 19.8 & 0.2243 & 18.29 &  293 &  21 & 17 & U & 0.2 \\
15 39 22.61 & 34 41 43.4 & 0.91   & 22.44 &  397 & 102 & 31 & R & 0.1 \\
15 39 23.66 & 34 28 25.7 & 0.00   & 21.36 &  178 &  34 & 15 & R & 0.5 \\
15 39 24.70 & 34 11 46.3 & 0.82   & 21.47 & 3867 &  33 & 24 & U & 0.4 \\
15 39 24.72 & 34 17 28.3 & 0.54   & 20.68 &  100 &  31 & 18 & R & 0.4 \\
15 39 24.86 & 34 39 22.0 & 0.02   & 22.26 &  721 &  63 & 27 & R & 0.2 \\
15 39 25.46 & 34 38 16.8 & 0.0791\tablenotemark{b} & 16.48 &  537 & 113 & 14 & R & 0.6 \\
15 39 26.28 & 34 17 14.3 & 0.66   & 21.40 &  168 &  39 & 16 & R & 0.1 \\
15 39 28.10 & 34 17 35.5 & 0.40   & 21.82 &  106 &  19 & 15 & U & 0.7 \\
15 39 29.11 & 34 25 38.6 & 0.1356 & 17.80 &  210 &  42 & 13 & R & 0.1 \\
15 39 29.52 & 34 32 42.4 & 0.22   & 19.36 &  189 &  36 & 16 & R & 0.0 \\
15 39 29.59 & 34 17 32.3 & 0.04   & 23.18\tablenotemark{a} &  109 &  18 & 15 & U & 0.3 \\
15 39 29.59 & 34 29 46.0 & 0.2210 & 18.97 &  160 &  38 & 15 & R & 0.5 \\
15 39 29.83 & 34 26 11.4 & 0.52   & 21.66\tablenotemark{a} &   82 &  13 & 14 & U & 0.2 \\
15 39 30.86 & 34 14 26.9 & 0.69   & 22.63 &  101 &  24 & 18 & U & 0.5 \\
15 39 30.98 & 34 29 33.0 & 0.18   & 19.28 &  294 &  57 & 14 & R & 0.3 \\
15 39 32.33 & 34 21 15.5 & 0.18   & 20.08 &  109 &  14 & 13 & U & 0.4 \\
15 39 32.78 & 34 26 17.9 & 0.63   & 21.36 &  172 &  13 & 13 & U & 0.2 \\
15 39 33.02 & 34 22 25.0 & 0.2299 & 18.63 &  204 &  28 & 12 & R & 0.1 \\
15 39 33.12 & 34 20 12.8 & 0.48   & 20.21 &  116 &  36 & 13 & R & 0.6 \\
15 39 34.37 & 34 34 51.6 & 0.35   & 20.35 &  183 &  47 & 17 & R & 0.1 \\
15 39 35.11 & 34 37 21.7 & 0.58   & 20.81 &  136 &  29 & 21 & U & 0.5 \\
15 39 35.62 & 34 26 40.2 & 0.39   & 19.81 &  120 &  36 & 13 & R & 0.3 \\
15 39 36.07 & 34 33 52.2 & 1.01   & 21.11 &  232 &  53 & 16 & R & 0.8 \\
15 39 37.58 & 34 28 47.6 & 0.2331 & 18.02 &  104 &  14 & 13 & U & 0.1 \\
15 39 37.66 & 34 44 06.0 & 0.04   & 22.85 & 2097 &  71 & 43 & U & 0.1 \\
15 39 37.68 & 34 35 58.9 & 0.2258 & 19.29 &  104 &  26 & 19 & U & 0.4 \\
15 39 38.21 & 34 22 10.6 & 0.15   & 19.57 &  114 &  31 & 13 & R & 0.5 \\
15 39 38.42 & 34 23 40.2 & 0.2208 & 18.33 &  137 &  24 & 13 & R & 0.5 \\
15 39 38.47 & 34 21 43.2 & 0.16   & 20.02 &  163 &  27 & 15 & R & 0.3 \\
15 39 38.50 & 34 19 10.6 & 0.1271 & 18.24 &  286 &  43 & 13 & R & 0.4 \\
15 39 38.88 & 34 37 59.2 & 0.53   & 22.11 &  201 &  33 & 21 & U & 0.2 \\
15 39 39.50 & 34 20 17.9 & 0.35   & 18.85 &   97 &  15 & 14 & U & 0.5 \\
15 39 40.13 & 34 11 34.1 & 0.21   & 18.94 &  153 &  33 & 22 & U & 0.3 \\
15 39 40.49 & 34 25 37.6 & 1.01   & 23.44 &  137 &  13 & 13 & U & 0.5 \\
15 39 41.26 & 34 42 18.4 & 0.48   & 20.20 &  250 &  52 & 33 & U & 0.1 \\
15 39 41.35 & 34 27 23.8 & 0.46   & 21.34 &   76 &  26 & 13 & R & 0.1 \\
15 39 41.57 & 34 21 31.0 & 0.23   & 19.05 &  254 &  39 & 13 & R & 0.1 \\
15 39 43.70 & 34 25 37.6 & 0.22   & 19.74 &  222 &  39 & 13 & R & 0.6 \\
15 39 44.09 & 34 23 59.3 & 0.33   & 19.95 &  113 &  34 & 12 & R & 0.4 \\
15 39 44.14 & 34 35 03.8 & 0.5512\tablenotemark{b} & 19.87 & 1300 &  36 & 19 & R & 0.4 \\
15 39 44.21 & 34 20 46.0 & 0.3329 & 18.79 &  253 &  46 & 14 & R & 0.4 \\
15 39 44.71 & 34 28 23.2 & 0.63   & 22.14 &  133 &  27 & 15 & R & 0.3 \\
15 39 45.00 & 34 12 58.0 & 0.89   & 22.79 &  125 &  29 & 20 & U & 0.3 \\
15 39 45.46 & 34 35 07.4 & 0.58   & 19.76 &  301 &  24 & 19 & U & 0.4 \\
15 39 45.50 & 34 12 20.9 & 1.01   & 23.09 &  145 &  30 & 21 & U & 0.3 \\
15 39 45.74 & 34 16 25.3 & 0.2336 & 19.35 &  235 &  20 & 15 & U & 0.3 \\
15 39 45.79 & 34 12 15.8 & 0.76   & 22.52 &  263 &  31 & 21 & U & 0.2 \\
15 39 46.68 & 34 24 34.6 & 0.2274 & 18.03 &  178 &  23 & 12 & R & 0.2 \\
15 39 46.78 & 34 16 46.6 & 0.49   & 20.12 &  211 &  19 & 14 & U & 0.1 \\
15 39 46.85 & 34 16 31.8 & 0.40   & 20.87 &  156 &  37 & 15 & R & 0.4 \\
15 39 47.93 & 34 08 13.2 & 0.22   & 19.27 &  240 &  48 & 29 & U & 0.2 \\
15 39 48.60 & 34 21 44.6 & 0.2294 & 17.96 &  279 &  28 & 13 & R & 0.1 \\
15 39 49.15 & 34 28 08.0 & 0.80   & 21.84 &   92 &  14 & 13 & U & 0.5 \\
15 39 49.30 & 34 26 40.9 & 0.2297 & 18.56 &  654 &  27 & 13 & R & 0.3 \\
15 39 49.87 & 34 22 50.2 & 0.76   & 23.28 &   65 &  14 & 13 & U & 0.2 \\
15 39 49.87 & 34 46 56.6 & 0.25   & 20.19\tablenotemark{a} &  859 & 120 & 59 & U & 0.6 \\
15 39 51.31 & 34 24 00.7 & 0.49   & 19.83 &  428 &  13 & 13 & U & 0.2 \\
15 39 51.50 & 34 29 48.5 & 0.82   & 23.22 &  188 &  42 & 13 & R & 1.1 \\
15 39 51.53 & 34 37 23.2 & 0.20   & 19.83 &  195 &  51 & 21 & R & 0.2 \\
15 39 51.79 & 34 25 30.7 & 0.43   & 20.21 &  103 &  31 & 13 & R & 0.5 \\
15 39 52.61 & 34 36 59.8 & 0.32   & 20.20 &  240 &  30 & 20 & U & 0.3 \\
15 39 52.63 & 34 32 10.0 & 0.00   & 20.42 &  132 &  18 & 15 & U & 0.2 \\
15 39 54.89 & 34 17 08.2 & 0.2312\tablenotemark{c} & 18.98 &  409 &  38 & 15 & R & 0.5 \\
15 39 55.06 & 34 20 12.8 & 0.38   & 18.64 & 13065 & 25 & 15 & R & 0.4 \\
15 39 55.13 & 34 31 57.7 & 0.36   & 20.15 &   98 &  18 & 15 & U & 0.4 \\
15 39 55.44 & 34 17 16.1 & 0.00   & 21.20\tablenotemark{a} &   97 &  19 & 16 & U & 0.4 \\
15 39 55.68 & 34 24 20.9 & 0.43   & 21.64 &   97 &  14 & 14 & U & 0.2 \\
15 39 55.82 & 34 12 51.5 & 0.33   & 19.93 &  268 &  47 & 22 & R & 0.2 \\
15 39 56.54 & 34 15 06.1 & 0.1555 & 17.77 &  338 &  23 & 19 & U & 0.1 \\
15 39 56.57 & 34 19 17.8 & 0.44   & 20.66 &  204 &  35 & 15 & R & 0.6 \\
15 39 56.57 & 34 31 03.0 & 0.16   & 19.73 &  458 &  17 & 15 & U & 0.2 \\
15 39 56.88 & 34 29 55.0 & 0.63   & 22.31 &  113 &  16 & 15 & U & 0.3 \\
15 39 56.90 & 34 29 32.3 & 0.79   & 22.50 & 1558 &  16 & 15 & U & 0.1 \\
15 39 57.89 & 34 25 40.1 & 0.13   & 20.90 &  216 &  48 & 13 & R & 0.6 \\
15 39 58.78 & 34 30 05.4 & 0.53   & 21.43 &   80 &  17 & 15 & U & 0.3 \\
15 39 59.71 & 34 24 20.9 & 0.42   & 19.81 &  344 &  32 & 13 & R & 0.6 \\
15 40 02.04 & 34 39 58.3 & 0.64   & 21.62 &  930 &  43 & 28 & U & 0.2 \\
15 40 02.11 & 34 33 46.4 & 0.38   & 19.78 &  103 &  22 & 16 & U & 0.7 \\
15 40 02.76 & 34 30 35.3 & 1.01   & 22.70 &  334 &  19 & 15 & U & 0.3 \\
15 40 02.88 & 34 23 29.8 & 0.2344 & 18.56 &  282 &  61 & 13 & R & 0.3 \\
15 40 04.01 & 34 14 22.6 & 0.14   & 20.42 &  113 &  27 & 19 & U & 0.8 \\
15 40 04.10 & 34 23 40.9 & 0.32   & 19.93 &   91 &  32 & 14 & R & 0.3 \\
15 40 04.27 & 34 14 52.1 & 0.30   & 19.29 &  474 &  25 & 18 & U & 0.2 \\
15 40 04.32 & 34 24 59.4 & 0.3263 & 19.03 &  193 &  32 & 15 & R & 0.1 \\
15 40 04.85 & 34 17 01.7 & 0.30   & 20.06 &  166 &  21 & 17 & U & 0.3 \\
15 40 04.94 & 34 08 50.6 & 0.43   & 21.98 &  205 &  56 & 28 & R & 0.4 \\
15 40 07.15 & 34 41 12.8 & 1.01   & 23.79 &  375 &  53 & 30 & U & 0.5 \\
15 40 08.14 & 34 18 36.4 & 0.46   & 19.43 &  179 &  37 & 16 & R & 0.2 \\
15 40 09.02 & 34 31 28.2 & 0.43   & 19.08 &  249 &  21 & 16 & U & 0.2 \\
15 40 09.43 & 34 14 38.4 & 0.40   & 20.58 &  123 &  27 & 19 & U & 0.4 \\
15 40 09.96 & 34 12 53.6 & 0.0481 & 16.77 &  969 & 221 & 22 & R & 0.3 \\
15 40 10.63 & 34 37 53.8 & 0.05   & 18.33\tablenotemark{a} &  475 &  60 & 24 & R & 0.1 \\
15 40 10.94 & 34 15 13.3 & 0.2308 & 18.90 &  318 &  26 & 20 & U & 0.2 \\
15 40 12.55 & 34 21 40.7 & 0.77   & 23.48 &  257 &  18 & 15 & U & 0.3 \\
15 40 12.72 & 34 12 16.2 & 0.17   & 18.99 &  351 &  72 & 23 & R & 0.3 \\
15 40 16.20 & 34 18 24.1 & 0.14   & 20.05 &  452 &  54 & 18 & R & 0.2 \\
15 40 17.66 & 34 37 46.6 & 0.19   & 17.91 &  181 &  38 & 24 & U & 0.1 \\
15 40 20.62 & 34 08 17.2 & 0.44   & 20.01 &  378 &  98 & 33 & R & 0.9 \\
15 40 20.69 & 34 21 32.8 & 0.59   & 22.21 &  117 &  20 & 16 & U & 0.6 \\
15 40 21.72 & 34 21 30.6 & 0.60   & 22.58 &  128 &  21 & 16 & U & 0.4 \\
15 40 21.79 & 34 13 48.4 & 0.70   & 21.76 &  799 &  35 & 22 & U & 0.3 \\
15 40 22.20 & 34 33 57.2 & 0.57   & 21.03 &  185 &  46 & 19 & R & 0.2 \\
15 40 22.25 & 34 21 31.3 & 0.0796 & 18.12 &  179 &  39 & 16 & R & 0.3 \\
15 40 22.56 & 34 26 40.9 & 0.70   & 22.79 &  636 &  21 & 17 & U & 0.2 \\
15 40 23.30 & 34 33 57.2 & 0.2981 & 19.35 &  303 &  64 & 20 & R & 0.1 \\
15 40 24.74 & 34 27 23.0 & 0.0755 & 17.74 &  820 &  39 & 19 & R & 0.1 \\
15 40 26.57 & 34 26 38.0 & 0.90   & 22.35 &  109 &  23 & 18 & U & 1.2 \\
15 40 27.86 & 34 32 38.0 & 0.39   & 22.57 &  128 &  29 & 21 & U & 0.4 \\
15 40 29.47 & 34 34 07.7 & 0.62   & 23.04 &  410 &  40 & 24 & U & 0.1 \\
15 40 32.06 & 34 38 22.6 & 0.23   & 19.61\tablenotemark{a} &   389 &  51 & 31 & U & 0.4 \\
15 40 32.26 & 34 24 34.9 & 0.95   & 21.76 &  144 &  24 & 17 & U & 1.0 \\
15 40 33.24 & 34 33 24.1 & 0.25   & 18.82 &  233 &  40 & 23 & U & 0.3 \\
15 40 33.62 & 34 34 53.0 & 0.2975 & 18.10 &  160 &  45 & 25 & U & 1.0 \\
15 40 35.47 & 34 23 46.0 & 0.35   & 18.79 &  122 &  25 & 19 & U & 1.3 \\
15 40 35.69 & 34 26 03.5 & 0.58   & 23.10 & 1831 &  41 & 18 & R & 0.2 \\
15 40 41.18 & 34 19 48.4 & 0.60   & 21.67 & 5135 &  51 & 23 & R & 0.1 \\
15 40 41.38 & 34 23 02.8 & 0.28   & 19.05 &  284 &  30 & 21 & U & 0.3 \\
15 40 41.81 & 34 18 37.1 & 0.50   & 19.91 &10298\tablenotemark{d} & 164 & 25 & R & 0.6 \\
15 40 53.23 & 34 32 31.2 & 0.45   & 22.36 &  251 &  75 & 34 & R & 1.0 \\
15 40 54.43 & 34 16 43.7 & 0.18   & 19.29 &  295 &  54 & 32 & U & 0.5 \\
15 40 55.13 & 34 30 15.5 & 0.1942\tablenotemark{b} & 16.49 & 15741 & 86 & 34 & R & 0.1 \\
15 41 03.98 & 34 26 05.6 & 0.60   & 20.56 &  328 &  54 & 30 & U & 0.4 \\
15 41 08.38 & 34 35 23.6 & 0.11   & 17.12 &  392 &  95 & 50 & U & 1.1 \\
15 41 09.60 & 34 29 40.2 & 0.74   & 21.83 &  305 &  72 & 38 & U & 0.5 \\
15 41 09.89 & 34 18 31.7 & 1.01   & 20.71 & 1019 & 184 & 41 & R & 0.5 \\
15 41 21.19 & 34 31 08.8 & 0.77   & 23.53 & 1416 & 191 & 56 & R & 0.4 \\
15 41 25.58 & 34 28 30.4 & 0.1842 & 17.47 &10215\tablenotemark{e} & 431 & 57 & R & 1.4 \\
15 41 28.73 & 34 26 48.8 & 0.47   & 21.58 &  477 & 151 & 59 & R & 0.6 \\
\enddata

\tablenotetext{a}{Classified as star in SDSS photometric data.}
\tablenotetext{b}{Redshift from SDSS DR4.}
\tablenotetext{c}{Redshift from \citet{morrison2003}}
\tablenotetext{d}{Extended radio galaxy with morphology of a compact double, with flux calculated by TVSTAT. The separation listed is for the midpoint of the two radio lobes relative to the optical position. There is additional diffuse to the southeast, possibly associated with the galaxy. Including this emission the total flux is 19.8 mJy with an error of 0.5 mJy. The galaxy and its radio emission are depicted in Figure \ref{fig-J154041}.}
\tablenotetext{e}{Extended radio source, TVSTAT}

\end{deluxetable}

\begin{deluxetable}{l l c r c r c}
\tablecolumns{7}
\tablecaption{Spectroscopic Classifications of Cluster Members\label{tbl-spec}}
\tablewidth{0pt}
\tablehead{
\colhead{RA(J2000)} & \colhead{Dec(J2000)} & \colhead{Class} & 
\colhead{SN} & \colhead{Color} & \colhead{Dist} & \colhead{Radio}
}
\startdata
15 37 53.32 & +34 36 34.2 & e(c) & 12.5 & R & 5.43 & N \\ 
15 37 57.38 & +34 42 34.9 & e(c) & 11.6 & R & 5.99 & O \\
15 38 14.68 & +34 39 45.3 & k    & 11.1 & R & 5.01 & O \\
15 38 45.05 & +34 32 16.2 & e(a) & 13.8 & B & 2.95 & N \\ 
15 38 50.00 & +34 32 36.2 & e(c) & 10.0 & B & 2.81 & N \\
15 38 54.62 & +34 28 24.6 & e(c) & 11.8 & B & 2.20 & N \\
15 38 55.10 & +34 24 30.4 & k    & 11.1 & R & 2.06 & N \\
15 39 01.75 & +34 26 54.0 & k    & 11.8 & R & 1.80 & N \\
15 39 05.82 & +34 31 34.1 & k    & 12.9 & R & 2.11 & N \\
15 39 05.98 & +34 33 06.4 & k    & 14.1 & R & 2.35 & N \\
15 39 07.54 & +34 35 47.9 & k    &  9.9 & R & 2.77 & N \\
15 39 11.06 & +34 27 43.8 & k    & 12.2 & R & 1.46 & N \\
15 39 13.32 & +34 31 50.8 & k    & 16.1 & R & 1.92 & N \\ 
15 39 14.83 & +34 33 25.2 & k    & 12.7 & R & 2.16 & N \\
15 39 16.78 & +34 28 51.0 & k    & 11.6 & R & 1.36 & D \\
15 39 17.43 & +34 21 15.9 & k    & 12.2 & R & 1.34 & N \\
15 39 20.07 & +34 14 12.8 & k    &  8.9 & R & 2.54 & N \\ 
15 39 20.84 & +34 26 01.4 & k    & 14.0 & R & 0.92 & N \\
15 39 21.52 & +34 33 19.9 & e(a) & 13.1 & B & 2.00 & D \\
15 39 24.86 & +34 34 44.1 & k    & 15.4 & R & 2.22 & N \\
15 39 28.38 & +34 25 19.9 & k    & 13.3 & R & 0.56 & N \\
15 39 28.42 & +34 27 19.2 & k    & 15.1 & R & 0.74 & N \\
15 39 29.46 & +34 33 46.8 & k    &  9.4 & R & 1.96 & N \\
15 39 29.50 & +34 31 11.7 & k+a  &  9.7 & R & 1.43 & N \\
15 39 29.59 & +34 29 46.2 & e(a) & 12.4 & B & 1.14 & D \\
15 39 31.99 & +34 53 50.5 & e(a) & 11.7 & B & 6.27 & O \\
15 39 33.01 & +34 22 25.1 & e(c) &  9.8 & R & 0.68 & D \\
15 39 34.31 & +34 26 13.2 & k    & 13.4 & R & 0.39 & N \\
15 39 34.34 & +34 30 17.4 & k    & 10.9 & R & 1.17 & N \\
15 39 36.81 & +34 25 39.7 & k    & 10.7 & R & 0.22 & N \\
15 39 37.22 & +34 29 56.7 & k    & 12.4 & R & 1.07 & N \\
15 39 37.59 & +34 28 47.9 & k    & 12.4 & R & 0.83 & D \\
15 39 37.71 & +34 27 04.0 & k    & 12.9 & R & 0.46 & N \\
15 39 37.68 & +34 35 58.9 & e(a) & 10.0 & B & 2.38 & D \\
15 39 37.93 & +34 28 31.5 & k    & 13.0 & R & 0.76 & N \\
15 39 38.37 & +34 27 31.0 & k    & 10.2 & R & 0.54 & N \\
15 39 38.44 & +34 23 40.9 & e(a) & 13.4 & B & 0.32 & D \\
15 39 38.75 & +34 26 39.0 & k+a  & 10.7 & B & 0.36 & N \\ 
15 39 39.40 & +34 25 14.1 & k    & 12.7 & R & 0.08 & N \\
15 39 40.50 & +34 25 27.8 & k    & 13.4 & R & 0.09 & N \\
15 39 40.98 & +34 22 49.6 & k    & 10.9 & R & 0.49 & N \\
15 39 41.32 & +34 24 35.1 & k    & 10.0 & R & 0.11 & N \\
15 39 41.67 & +34 32 11.9 & k    & 11.8 & R & 1.55 & N \\
15 39 41.79 & +34 24 43.4 & k    & 14.6 & R & 0.08 & N \\
15 39 41.91 & +34 24 20.1 & k+a  & 18.3 & B & 0.17 & N \\ 
15 39 42.01 & +34 26 30.5 & k    & 14.2 & R & 0.32 & N \\
15 39 42.82 & +34 29 04.8 & k    & 10.3 & R & 0.88 & N \\
15 39 43.52 & +34 21 45.5 & k    & 11.0 & R & 0.73 & N \\
15 39 44.31 & +34 16 49.9 & k    & 15.1 & R & 1.80 & N \\
15 39 45.73 & +34 16 25.8 & e(n) & 10.4 & R & 1.89 & D \\
15 39 45.74 & +34 26 57.5 & k    & 11.3 & R & 0.47 & N \\
15 39 46.15 & +34 16 05.4 & k    & 11.1 & R & 1.97 & N \\
15 39 46.58 & +34 19 38.1 & k    & 12.3 & R & 1.21 & N \\
15 39 46.69 & +34 24 35.1 & k    & 14.9 & R & 0.28 & D \\
15 39 47.09 & +34 27 37.5 & k    & 13.1 & B & 0.62 & N \\ 
15 39 47.29 & +34 25 10.5 & k    &  9.9 & R & 0.29 & N \\
15 39 47.70 & +34 30 07.8 & k    &  9.9 & B & 1.14 & N \\ 
15 39 47.74 & +34 21 23.4 & k    &  8.5 & R & 0.86 & N \\ 
15 39 48.50 & +34 22 36.1 & k    & 12.3 & R & 0.64 & N \\
15 39 48.61 & +34 21 45.1 & e(a) & 16.5 & B & 0.80 & D \\ 
15 39 48.71 & +34 23 34.9 & k    & 12.9 & R & 0.48 & N \\
15 39 49.30 & +34 26 41.4 & e(a) & 12.4 & B & 0.52 & D \\
15 39 49.77 & +34 23 14.3 & k    & 14.2 & R & 0.56 & N \\
15 39 50.48 & +34 31 48.4 & e(c) &  9.4 & B & 1.53 & N \\
15 39 51.15 & +34 28 39.3 & e(a) & 12.6 & R & 0.91 & N \\
15 39 51.26 & +34 36 11.2 & k    &  9.8 & R & 2.46 & N \\
15 39 52.50 & +34 30 26.5 & e(a) & 13.3 & B & 1.28 & N \\
15 39 52.93 & +34 16 54.8 & k    & 11.8 & R & 1.85 & N \\
15 39 52.96 & +34 27 48.0 & e(c) & 11.0 & R & 0.80 & N \\
15 39 53.19 & +34 24 03.9 & k    & 12.5 & R & 0.59 & N \\
15 39 54.11 & +34 22 04.1 & k    &  7.5 & R & 0.88 & N \\ 
15 39 54.35 & +34 32 28.2 & k+a  & 12.4 & R & 1.72 & N \\ 
15 39 54.67 & +34 24 56.7 & k    & 15.6 & R & 0.62 & N \\
15 39 57.59 & +34 30 19.0 & k    &  9.2 & R & 1.37 & N \\
15 39 59.44 & +34 17 13.1 & k    & 13.7 & R & 1.90 & N \\
15 39 59.96 & +34 24 27.6 & k    & 10.8 & R & 0.87 & N \\
15 40 00.45 & +34 25 01.5 & k    & 10.6 & R & 0.88 & N \\
15 40 01.94 & +34 18 27.1 & e(a) & 12.1 & B & 1.72 & N \\
15 40 02.34 & +34 26 25.7 & k+a  &  9.4 & R & 1.01 & N \\ 
15 40 02.88 & +34 23 30.6 & e(c) & 12.4 & B & 1.04 & D \\ 
15 40 03.12 & +34 34 33.2 & k    &  9.8 & R & 2.29 & N \\
15 40 05.60 & +34 15 20.2 & k    & 10.2 & R & 2.39 & N \\
15 40 07.89 & +34 29 54.0 & k    &  9.9 & R & 1.60 & N \\
15 40 08.96 & +34 16 03.7 & e(a) &  9.3 & B & 2.33 & N \\ 
15 40 10.94 & +34 15 13.3 & e(c) & 12.2 & R & 2.53 & D \\ 
15 40 21.39 & +34 22 10.5 & k    & 13.7 & R & 1.92 & N \\
15 40 23.44 & +34 17 26.5 & e(c) &  7.5 & B & 2.53 & N \\ 
15 40 27.19 & +34 16 34.5 & e(c) & 11.6 & B & 2.78 & N \\
15 40 27.94 & +34 13 37.7 & e(a) & 10.1 & R & 3.27 & N \\
15 40 29.18 & +34 27 22.0 & e(c) & 11.1 & B & 2.22 & N \\
15 40 30.92 & +34 30 50.4 & k    & 12.1 & R & 2.57 & N \\
15 40 31.97 & +34 29 19.0 & k    &  8.8 & R & 2.47 & N \\
15 40 36.07 & +34 04 52.3 & k    & 16.8 & R & 5.05 & N \\  
15 40 52.10 & +34 27 37.8 & k    & 16.0 & R & 3.24 & N \\
15 41 45.24 & +34 21 39.2 & e(a) & 10.2 & B & 5.63 & N \\
\enddata

\tablecomments{SN is the signal-to-noise ratio per resolution element in the vicinity of \hd. Color is simply R for consistent with red sequence, B if blue in BO84 sense (i.e., $\Delta(g-r)\leq-0.2$). Distance is in Mpc from cluster center. Radio is N for non-detected, D for detected and in Table \ref{tbl-props}, and O for outside surveyed area.}

\end{deluxetable}

\begin{deluxetable}{l l r r r r r r c}
\tablecolumns{9}
\tablecaption{Properties of Cluster Radio Sources with Optical Spectra\label{tbl-props}}
\tablewidth{0pt}
\tablehead{
\colhead{RA(J2000)} & \colhead{Dec(J2000)} & \colhead{$z$} & \colhead{$M_R$} & 
\colhead{$\Delta (g - r)$} & \colhead{log($L_{1.4}$)} & \colhead{SFR} & 
\colhead{r} & \colhead{Class}
}
\startdata
15 39 16.78 & 34 28 50.9 & 0.2293 & -22.11 &  0.04 & 22.34 & 12.9\tablenotemark{a} & 1.36 & k    \\
15 39 21.50 & 34 33 19.8 & 0.2243 & -22.39 & -0.56 & 22.62 & 24.8 & 2.00 & e(a) \\
15 39 29.59 & 34 29 46.0 & 0.2210 & -21.64 & -0.58 & 22.36 & 13.5 & 1.14 & e(a) \\
15 39 33.02 & 34 22 25.0 & 0.2299 & -22.17 & -0.05 & 22.47 & 17.3 & 0.68 & e(c) \\
15 39 37.58 & 34 28 47.6 & 0.2331 & -22.74 & -0.06 & 22.17 &  8.8\tablenotemark{a} & 0.83 & k    \\
15 39 37.68 & 34 35 58.9 & 0.2258 & -21.37 & -0.47 & 22.17 &  8.8 & 2.38 & e(a)\tablenotemark{b} \\
15 39 38.42 & 34 23 40.2 & 0.2208 & -22.39 & -0.31 & 22.29 & 11.6 & 0.33 & e(a)  \\
15 39 45.74 & 34 16 25.3 & 0.2336 & -21.30 &  0.00 & 22.53 & 19.9\tablenotemark{a} & 1.89 & e(n)\tablenotemark{c} \\
15 39 46.68 & 34 24 34.6 & 0.2274 & -22.78 &  0.05 & 22.41 & 15.1\tablenotemark{a} & 0.28 & k    \\
15 39 48.60 & 34 21 44.6 & 0.2294 & -22.77 & -0.23 & 22.60 & 23.6 & 0.80 & e(a) \\
15 39 49.30 & 34 26 40.9 & 0.2297 & -22.19 & -0.35 & 22.97 & 55.3 & 0.51 & e(a) \\
15 39 54.89 & 34 17 08.2 & 0.2312\tablenotemark{d} & -21.79 &  0.34 & 22.77 & 34.6\tablenotemark{a} & 1.83 & \nodata \\
15 40 02.88 & 34 23 29.8 & 0.2344 & -22.14 & -0.54 & 22.61 & 23.9 & 1.04 & e(c) \\
15 40 10.94 & 34 15 13.3 & 0.2308 & -21.82 & -0.10 & 22.66 & 26.9 & 2.53 & e(c) \\
\enddata

\tablenotetext{a}{Based on the color and/or optical spectrum, the galaxy is likely to be an AGN and hence the SFR is an upper limit due to inclusion of radio emission associated with the active nucleus.}
\tablenotetext{b}{Noisy spectrum, so \hd{} absorption may be insufficient for e(a) classification.}
\tablenotetext{c}{Seyfert 2 spectrum.}
\tablenotetext{d}{Velocity reported in \citet{morrison2003}.}

\tablecomments{Apparent $R$ magnitudes were estimated from the SDSS $u~g~r~i~z$ magnitudes using the observed redshifts and {\ttfamily kcorrect} software \citep{blanton2003}, with the absolute $R$ magnitudes based on $D_L = 1129$ Mpc, $z=0.2287$, and $A_R=0.07$. $\Delta (g - r)$ is color relative to the cluster CM relation. Values less than -0.2 are thus considered blue in the Butcher-Oemler sense. The radio luminosities were calculated assuming a spectral index of 0.7.}

\end{deluxetable}

\begin{deluxetable}{l l r r r r r r}
\tablecolumns{8}
\tablecaption{Properties of Radio Sources with Photometric Redshifts Consistent with Cluster Membership\label{tbl-pzprops}}
\tablewidth{0pt}
\tablehead{
\colhead{RA(J2000)} & \colhead{Dec(J2000)} & \colhead{$z_{phot}$} & 
\colhead{$M_R$} & \colhead{$\Delta (g - r)$} & \colhead{log($L_{1.4}$)} &
\colhead{SFR} & \colhead{r}
}
\startdata
15 39 05.83 & 34 28 35.4 & 0.18 & -20.9 & -0.44 & 22.24 & 10.2 & 1.75 \\
15 39 29.52 & 34 32 42.4 & 0.22 & -21.4 & -0.31 & 22.43 & 16.0 & 1.74 \\
15 39 30.98 & 34 29 33.0 & 0.18 & -21.4 & -0.54 & 22.62 & 24.9 & 1.07 \\
15 39 32.33 & 34 21 15.5 & 0.18 & -20.7 & -0.10 & 22.19 &  9.2 & 0.92 \\
15 39 38.21 & 34 22 10.6 & 0.15 & -21.1 & -0.39 & 22.21 &  9.6 & 0.64 \\
15 39 38.47 & 34 21 43.2 & 0.16 & -20.6 & -0.43 & 22.37 & 13.8 & 0.74 \\
15 39 39.50 & 34 20 17.9 & 0.35 & -22.1 &  0.15 & 22.14 &  8.2 & 1.04 \\
15 39 41.57 & 34 21 31.0 & 0.23 & -21.8 &  0.01 & 22.56 & 21.5 & 0.77 \\
15 39 43.70 & 34 25 37.6 & 0.22 & -21.0 & -0.43 & 22.50 & 18.8 & 0.17 \\
15 39 44.09 & 34 23 59.3 & 0.33 & -20.7 & -0.28 & 22.21 &  9.6 & 0.27 \\
15 39 56.57 & 34 31 03.0 & 0.16 & -21.0 & -0.18 & 22.82 & 38.8 & 1.48 \\
15 40 04.10 & 34 23 40.9 & 0.32 & -20.9 & -0.13 & 22.12 &  7.7 & 1.08 \\
\enddata

\tablecomments{The photometric redshifts and apparent $R$ magnitudes were determined using the SDSS $u~g~r~i~z$ magnitudes and {\ttfamily kcorrect} software \citep{blanton2003}. The absolute $R$ magnitudes and radio luminosities were then calculated as in Table \ref{tbl-props} assuming the galaxies were cluster members with $z=0.2287$ and $D_L = 1129$ Mpc. $\Delta (g - r)$ is the color relative to the cluster CM relation. Values less than -0.2 are thus considered blue in the Butcher-Oemler sense.}

\end{deluxetable}

\begin{deluxetable}{l l r r r r c}
\tablecolumns{7}
\tablecaption{Properties of Foreground/Background Radio Sources with Optical Spectra\label{tbl-bgprops}}
\tablewidth{0pt}
\tablehead{
\colhead{RA(J2000)} & \colhead{Dec(J2000)} & \colhead{$z$} & \colhead{$M_R$} & 
\colhead{$B - V$} & \colhead{log($L_{1.4}$)} & \colhead{Class} 
}
\startdata
15 38 40.53 & 34 23 44.9 & 0.8718 & -25.12 & 0.12 & 23.76 & QSO\tablenotemark{a} \\ 
15 38 45.10 & 34 35 04.9 & 0.1292 & -20.90 & 0.84 & 21.90 & e(c)\tablenotemark{b,c} \\
15 38 52.44 & 34 31 43.0 & 0.1365 & -21.52 & 0.84 & 21.73 & k\tablenotemark{d} \\ 
15 39 06.67 & 34 39 28.8 & 0.1281 & -21.85 & 0.36 & 22.22 & e(b)\tablenotemark{a,e} \\ 
15 39 08.62 & 34 24 57.2 & 0.1487 & -21.29 & 0.39 & 22.09 & e(c)\tablenotemark{f} \\ 
15 39 11.38 & 34 17 51.7 & 0.2877 & -22.07 & 0.55 & 22.69 & e(c) \\ 
15 39 16.51 & 34 30 09.7 & 0.1557 & -20.62 & 0.38 & 22.20 & e(b)\tablenotemark{b} \\ 
15 39 25.46 & 34 38 16.8 & 0.0791 & -21.44 & 0.31 & 21.90 & e(a)\tablenotemark{a} \\ 
15 39 29.11 & 34 25 38.6 & 0.1356 & -21.55 & 0.58 & 21.99 & e(a)\tablenotemark{b} \\
15 39 38.50 & 34 19 10.6 & 0.1271 & -20.92 & 0.57 & 22.06 & e(c)\tablenotemark{b} \\ 
15 39 44.14 & 34 35 03.8 & 0.5512 & -23.57 & 0.49 & 24.14 & e(n)\tablenotemark{a,g} \\ 
15 39 44.21 & 34 20 46.0 & 0.3329 & -22.80 & 0.49 & 22.92 & e(c) \\ 
15 39 56.54 & 34 15 06.1 & 0.1555 & -21.94 & 0.69 & 22.32 & e(c) \\ 
15 40 04.32 & 34 24 59.4 & 0.3263 & -22.33 & 0.30 & 22.79 & e(b) \\ 
15 40 09.96 & 34 12 53.6 & 0.0481 & -20.05 & 0.51 & 21.71 & e(c)\tablenotemark{b,h} \\ 
15 40 22.25 & 34 21 31.3 & 0.0796 & -19.93 & 0.70 & 21.43 & e(a)\tablenotemark{b} \\ 
15 40 23.30 & 34 33 57.2 & 0.2981 & -22.18 & 0.73 & 22.89 & e(c)\tablenotemark{i} \\ 
15 40 24.74 & 34 27 23.0 & 0.0755 & -20.23 & 0.83 & 22.04 & e(c)\tablenotemark{b} \\ 
15 40 33.62 & 34 34 53.0 & 0.2975 & -23.28 & 0.72 & 22.62 & k \\ 
15 40 55.13 & 34 30 15.5 & 0.1942 & -23.83 & 0.81 & 24.20 & k\tablenotemark{a,j} \\
15 41 25.58 & 34 28 30.4 & 0.1842 & -22.69 & 0.82 & 23.96 & k \\ 
\enddata

\tablenotetext{a}{Spectrum from SDSS DR4.}
\tablenotetext{b}{Spectral wavelength coverage does not include \oii. The classification is based on \ha{} and other lines.}
\tablenotetext{c}{Possibly a very weak AGN on the basis of \nii{} relative to \ha, detected \oiii{} relative to non-detected \hb, and red rest frame color.}
\tablenotetext{d}{\oii{} and \ha{} outside of wavelength coverage, but no other emission lines observed. Fairly strong 4000$\mbox{\AA}$ break and red color conducive to k classification.}
\tablenotetext{e}{Formally an e(c) on basis of EW(\oii)$ < 40$, but EW(\ha+\nii)$ > 100$ warrants classification as e(b).}
\tablenotetext{f}{\hb{} is present but contaminated by telluric 5577 line, \oii{} and \ha{} outside wavelength coverage. \oiii{} emission is very slight and the rest frame color is blue, so tentative classification of e(c) is assigned.}
\tablenotetext{g}{Narrow line AGN, QSO in SDSS.}
\tablenotetext{h}{Spectral wavelength coverage does not include \hd. The classification is based on other Balmer lines.}
\tablenotetext{i}{Possibly a weak AGN, as the spectrum includes weak \oii{} and  \oiii{}, but no \hb. Similarly, the rest frame color is moderately red.}
\tablenotetext{j}{Plus detected \nii{} and \sii.}

\tablecomments{Absolute $R$ magnitudes and $B - V$ colors were estimated from the SDSS $u~g~r~i~z$ magnitudes using the observed redshifts and {\ttfamily kcorrect} software \citep{blanton2003}, with the absolute $R$ magnitudes also including $A_R=0.07$. The radio luminosities were calculated assuming a spectral index of 0.7.}

\end{deluxetable}

\end{document}